\documentclass[a4paper,12pt]{article}
\usepackage[top=3cm, bottom=3cm, left=3cm, right=3cm]{geometry} 
\usepackage{color}
\usepackage{rotating}
\usepackage{graphicx}
\usepackage{amsmath}
\usepackage{amssymb}
\usepackage{float}
\usepackage{subfig}
\usepackage{subfloat}
\usepackage{hyperref}
\def\displayandname#1{\rlap{$\displaystyle\csname #1\endcsname$}%
                      \qquad \texttt{\char92 #1}}

\begin{document}
\title{\bf{Optimisation and Characterisation of Glass RPC for India-based Neutrino Observatory Detectors}}

\author{R. Kanishka$^\alpha$\thanks{email:
kanishka.rawat.phy@gmail.com}~ \\
  Vipin Bhatnagar$^\beta$, D. Indumathi$^\gamma$ \\
  {\it $^\alpha$APND, Saha Institute of Nuclear Physics, Kolkata 700064, India}\\
{\it $^\beta$Physics Department, Panjab University, 
Sector 14, Chandigarh 160 014, India} \\
{\it $^\gamma$The Institute of Mathematical Sciences,
CIT Campus, Chennai 600 113, India} \\
}
\maketitle

\begin{abstract}
{The proposed magnetised Iron CALorimeter detector (ICAL) to be built in the India-based Neutrino Observatory (INO) laboratory aims to detect atmospheric muon neutrinos. In order to achieve improved physics results, the constituent components of the detector must be fully understood by proper characterisation and optimisation of various parameters. Resistive Plate Chambers (RPCs) are the active detector elements in the ICAL detector and can be made of glass or bakelite. The number of RPCs required for this detector are very large so a detailed study is necessary to establish the characterisation and optimisation of these RPCs. These detectors once installed will be taking data for 15-20 years. In this paper, we report the selection criteria of the glass electrodes procured from Indian manufacturers. Based on the factors that deteriorate the quality of glass the choice of electrode is made. The glass characterisation studies include UV-VIS transmission for optical properties, SEM, AFM for surface properties, WD-XRF, PIXE for determining the composition of glass samples and electrical properties. Based on these techniques a procedure is adopted to arrive at the best glass sample. We have done a second order check on the quality of the fabricated glass RPCs. In this regard, the efficiency and cross-talk of RPCs were measured. Results from Asahi and Saint Gobain glass RPCs came out to be the best.}

\end{abstract}

\newpage

\section{Introduction}

Neutrinos are one of the fundamental particles, occur in three flavours mainly; electron neutrino, muon neutrino and tau neutrino. While propagating, a neutrino could change its flavour and hence oscillates from one flavour to other \cite{kajita}. This change of flavour is due to the mass possessed by them. Many experiments around the globe are going on to measure various neutrino properties. On the similar lines ``India-based Neutrino Observatory'' (INO) \cite{ino:2006} with its magnetised detector Iron CALorimeter (ICAL) will be able to contribute to one such problem like mass hierarchy \cite{white} besides complementing the long baseline neutrino experiments data. The INO is the proposed underground facility to be built in Theni at Bodi west hills of south India. One of the experiments at the INO site for studying atmospheric neutrinos \cite{indu:2006, physics, physics1} will be ICAL. The ICAL detector comprises of three identical modules each having dimension of 16 m $\times$ 16 m $\times$ 14.45 m. The total number of layers in ICAL will be 150 with 5.6 cm thick iron plates interleaved by Resistive Plate Chambers (RPCs) \cite{satya}. The magnetic field of 1.3 T is applied in ICAL which is generated by passing current through copper coils, that pass through coil slots in the plates and is distributed non-uniformly, dividing the whole ICAL (each module) in three main regions. The ICAL will have good energy, direction resolution and good reconstruction and charge identification efficiencies \cite{central}. The ICAL being the target for neutrinos to interact and produce muons and hadrons \cite{hadrons} so we need to have an efficient tracking in the detector by using Resistive Plate Chambers (RPCs). These muons leave a track \cite{kolahal} that can be captured in one such detector called ``Resistive Plate Chambers'' (RPCs) \cite{Santonico, Santonico1}. So understanding of the RPCs in terms of operational characteristics needs to be done to get the maximum efficiency. In the next section we give the details of the RPC detectors.

\section{Resistive Plate Chambers}
\label{rpcs_paper}

\begin{figure}[htbp]
\renewcommand{\figurename}{Fig.}
\centering
\includegraphics[width=0.48\textwidth]{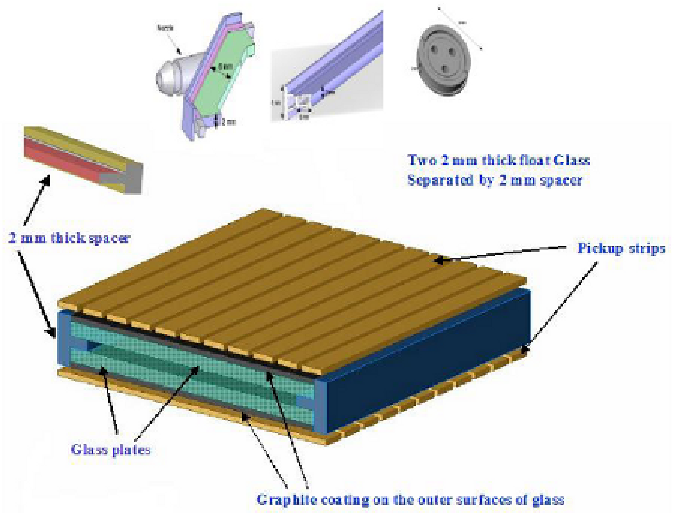}
\caption{A schematic of RPC detector.}
\label{rpc_view}
\end{figure}

The RPC \cite{satya} is a low cost detector which is being used in various experiments like BELLE \cite{belle}, CMS \cite{cms} and would be used in the near future experiments ICAL at INO. Fig.~\ref{rpc_view} shows schematic of RPC detector. 

The RPCs are the parallel plate gas detectors built using electrodes of high bulk resistivity such as glass or Bakelite. So, glass is the most important component in the RPC as any damage to them can effect the properties of particles. Based on excellent position and time resolutions they have been used for charged particle track detection and time of flight experiments. The RPC is fabricated by assembling two glass or Bakelite plates having bulk resistivity of $10^{10} - 10^{12}$ $\Omega$-cm, forming gap, filled with a certain gas mixture. Across this gap a high voltage is applied. A thin layer of graphite is coated over the external surface of the electrodes to permit uniform application of the high voltage. The electrodes are kept apart by means of small cylindrical spacers having a diameter of around 10 mm and a bulk resistivity greater than $10^{13}$ $\Omega$-cm (refer Fig.~\ref{rpc_view}). A gas mixture \cite{kalmani:2009} could consist of {\it {Argon}} which acts as target for ionising particles while {\it {Isobutane}}, being an organic gas, helps to absorb the photons that result from recombination processes thus limiting the formation of secondary avalanches far from the primary ones. An electronegative gas like {\it {Freon (R134a)}} serves the purpose of limiting the amount of free charge in the gas and $SF_{6}$ acts as a quenching gas. Electric signals are induced on pick-up strips, oriented in orthogonal directions and are placed on the outer surfaces of the glass electrodes. These pick-up strips are used to read the signal generated by RPC.

\paragraph{Modes of Operation}
There are two modes of operation for RPC depending on the operating voltage and gas composition used. {\bf Avalanche Mode} \cite{carda1, carda} operates at a lower voltage and the gain factor is less than $10^{8}$. It occurs when the external field opposes the electric field of the ionising particles and the multiplication process stops after sometime. Then the charges drift towards the electrodes and are collected there. This mode is used in those experiments where the event rate is low. {\bf Streamer Mode} \cite{fonte, fonte1} occurs when the secondary ionisation continues until there is a breakdown of the gas and a continuous discharge takes place. This mode operates at a high voltage and the gain factor is greater than $10^{8}$. This mode is used in those experiments where the event rate is high. Either operated in avalanche or streamer mode it has been reported \cite{aging} that the efficiency of RPC deteriorates and the reason accounted for such a low efficiency is assigned for aging of glass based RPCs. 

\subsection{Aging of Resistive Plate Chambers}

To ensure RPCs working efficiently for relatively long time period (couple of years), we need to understand the aging process so as to minimise the factors responsible for that. There are many factors which could lead to aging effect. There can be different kind of aging effects like, aging of the materials irrespective of the working conditions, due to the integrated dissipated current inside the detector, and due to irradiation. One of the reasons for the aging is the possible contamination of the detector gas with impurities and moisture too. One more reason is that, electron-ion recombination that produces UV photons that causes damage to the electrodes. Also, the glass material itself having some internal impurities can deteriorate the surface. First two factors can be controlled but third one has to be taken care by choosing the best glass electrode. The choice of gases is also important to prevent aging. One of the plausible chemical reactions of Freon gas, producing Fluoride radicals reacting with moisture, causes damage to the electrodes due to the formation of Hydrogen Fluoride (HF) inside the gas gap. So in order to improve the stability of RPCs, the characterisation of electrode material (glass) is important. Now, we describe the characterisation of glass  electrode, in order to understand and minimise the effect of aging. 

\section{Characterisation of Glass Electrode}
\label{char_glass}

Glass electrode seems to be one of the crucial component and helpful in minimising the aging of RPCs if proper choice of electrode is made. So we have done detailed characterisation studies based on various techniques, for various glass samples procured from different manufacturers from the domestic market \cite{Czyrkowski, Meghna}.

The following properties contribute to the glass characterisation process. 

{\bf {Physical Properties}}: Knowing the mass, length and breadth, we measured the density of all the glass samples. We observed no significant difference in the density. The results are given in the Table~\ref{table1}.

\begin{table}[ht]

\centering 
\begin{tabular}{|c c c c|} 
\hline 
Samples & Volume($cm^{3}$) & Weight(g) & Density($g/cm^{3}$\\ 
\hline                  
Asahi & 1.85 & 4.53 & 2.45 \\
Saint Gobain & 2.78 & 6.89 & 2.48 \\
Modi & 4.35 & 10.63 & 2.44 \\
\hline 
\end{tabular}
\caption{Density measurements of Asahi, Saint Gobain and Modi glass samples.} 
\label{table1} 
\end{table}

\begin{figure}[htbp]
\renewcommand{\figurename}{Fig.}
\centering
 \includegraphics[height=.25\textheight]{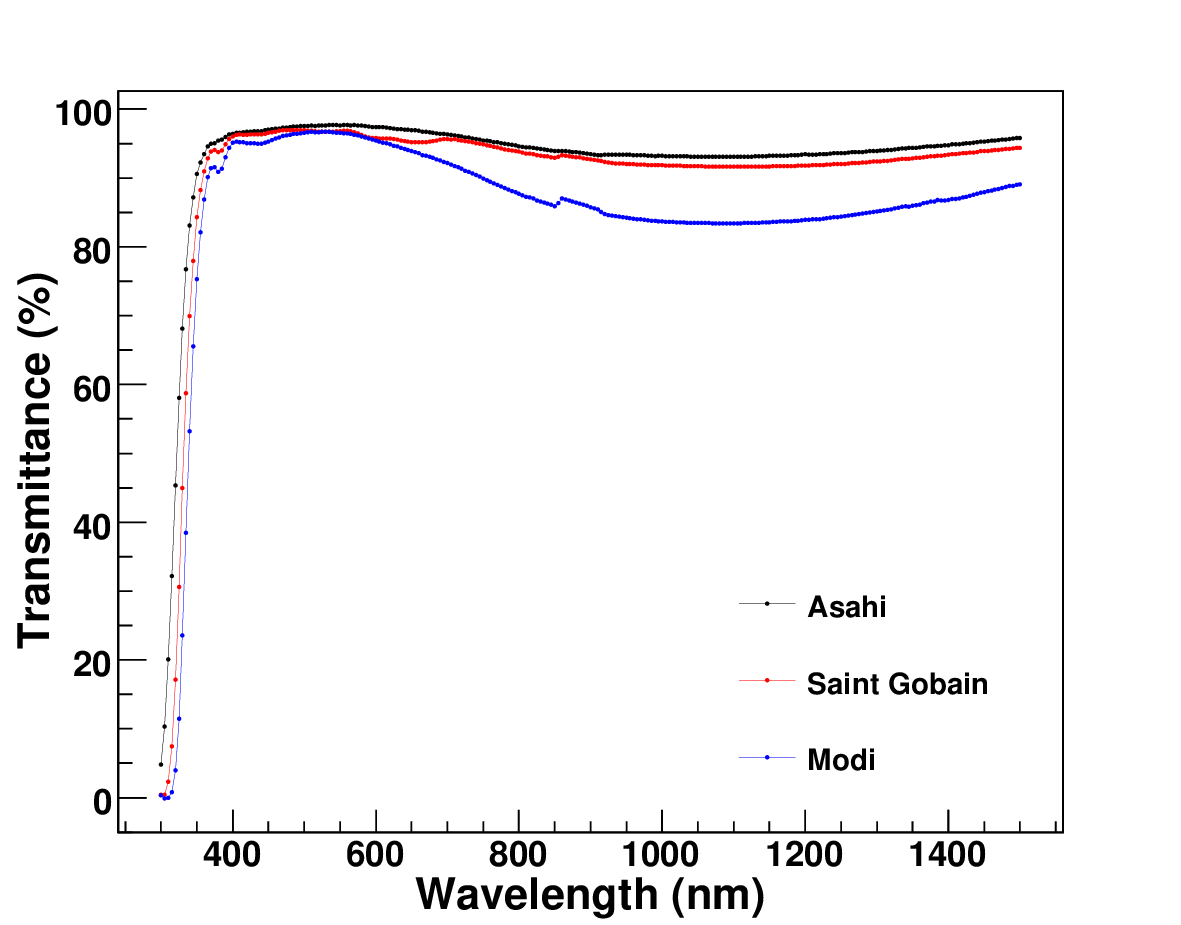}  
  \includegraphics[height=.25\textheight]{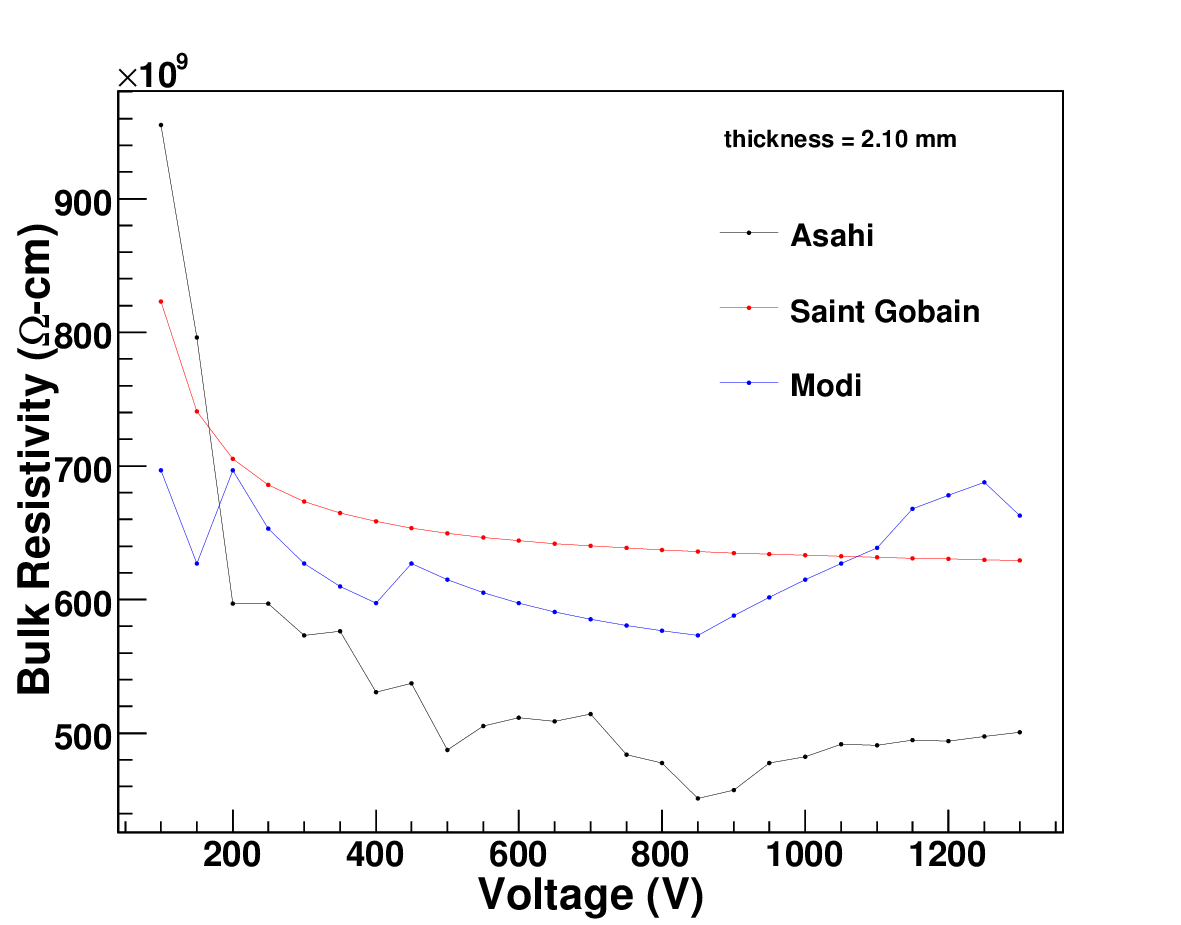}
  \caption{UV-VIS transmittance test (left) and bulk resistivity (right) of Asahi, Saint Gobain and Modi glass samples.}
\label{fig:bulk}
\end{figure}

{\bf {Optical Properties}}: Transmittance studies for various glass samples over the Ultraviolet to Visible light spectrum was carried out. This will indicate the general bulk quality of the glass and a measure of level of impurities in the glass. UV/VIS spectroscopy was used for the optical characteristics. Fig.~\ref{fig:bulk} (left) shows the optical transmittance for all the glass samples. Asahi and Saint Gobain glass shows better UV-VIS transmittance than Modi glass sample.

{\bf {Electrical Properties}}: The bulk resistivity of the glass samples was calculated using Two-Probe method. The Two Probe Method is one of the standard and most commonly used method for the measurement of resistivity of very high resistivity samples like sheets/films of polymers. Fig.~\ref{fig:bulk} (right) shows the bulk resistivity of Asahi, Saint Gobain and Modi (thickness 2.10 mm) and was found of the order of $10^{11}$ $\Omega$-cm.

\begin{figure}[htbp]
\renewcommand{\figurename}{Fig.}
\centering
  \includegraphics[height=.15\textheight]{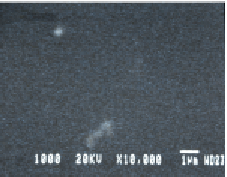}
\includegraphics[height=.15\textheight]{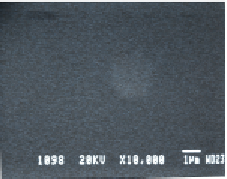}
\includegraphics[height=.15\textheight]{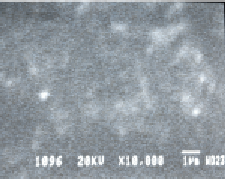}
  \caption{SEM scans of Asahi, Saint Gobain and Modi glass samples respectively.}
\label{fig:SEM}
\end{figure}

\begin{figure}[htbp]
\renewcommand{\figurename}{Fig.}
\centering
\includegraphics[height=.15\textheight]{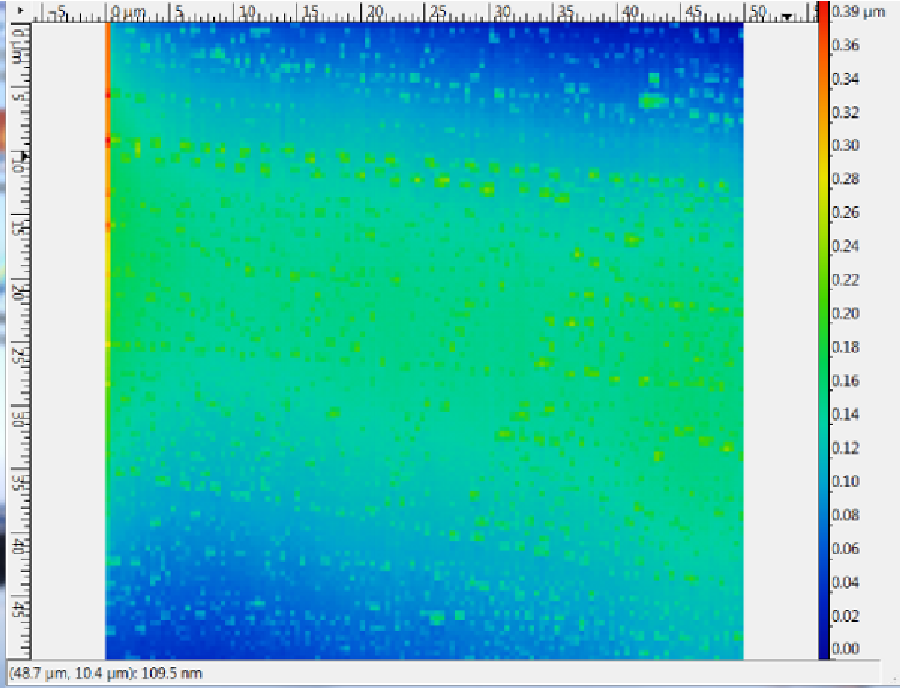}
\includegraphics[height=.15\textheight]{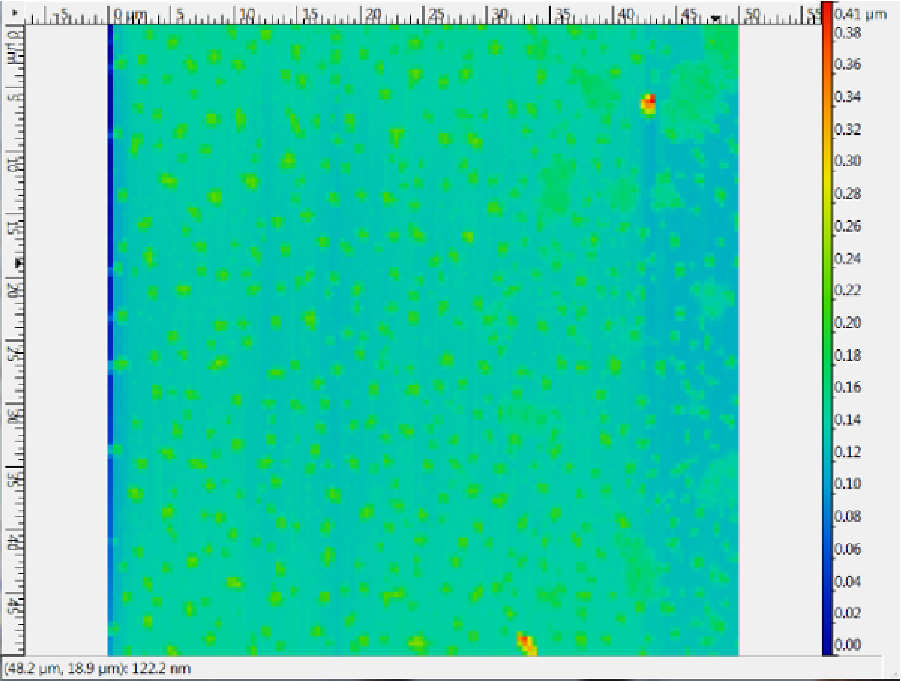}
  \includegraphics[height=.15\textheight]{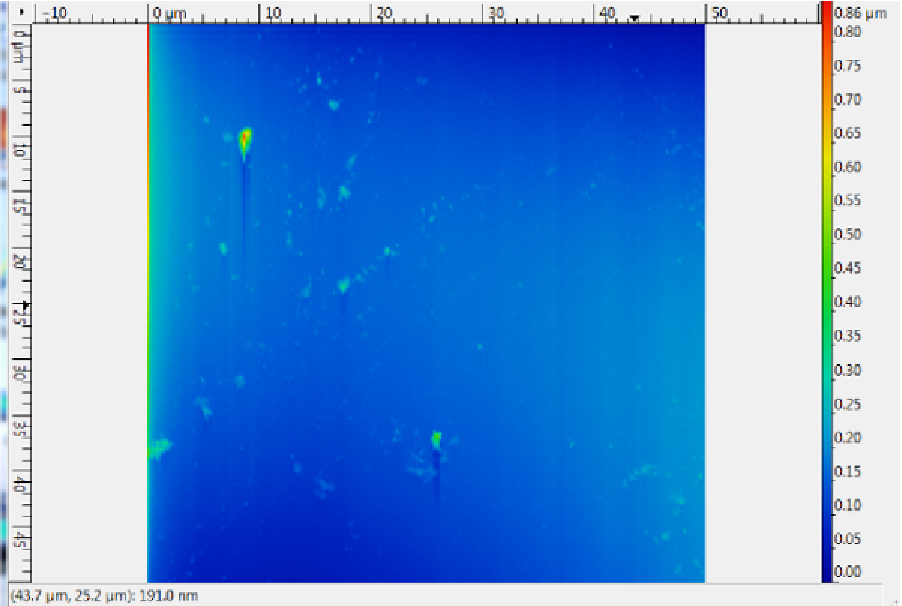}
  \caption{AFM scans of Asahi, Saint Gobain and Modi glass samples respectively.}
\label{fig:AFM}
\end{figure}

{\bf {Surface Properties}}: The surface quality of the electrode is crucial in reducing spontaneous discharges which might affect the rate capability of the detector. Atomic Force Microscopy (AFM) and Scanning Electron Microscopy (SEM) has been used to study and compare the surface quality of various glass samples. Fig.~\ref{fig:SEM} and ~\ref{fig:AFM} shows the SEM and AFM of all the glass samples. Asahi and Saint Gobain were better than Modi glass sample.

\begin{figure}[htbp]
\renewcommand{\figurename}{Fig.}
\centering
  \includegraphics[height=.16\textheight]{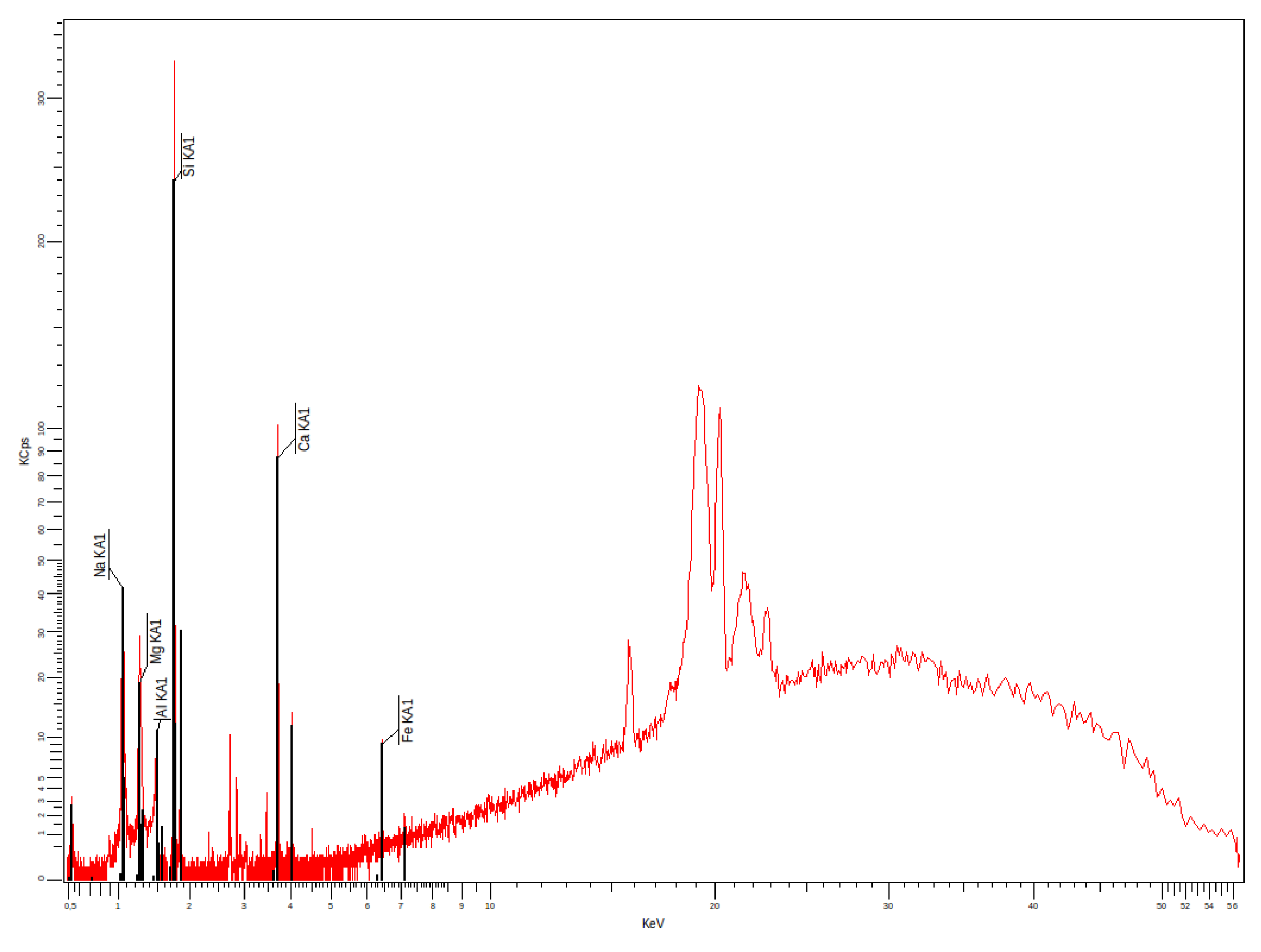}
\includegraphics[height=.16\textheight]{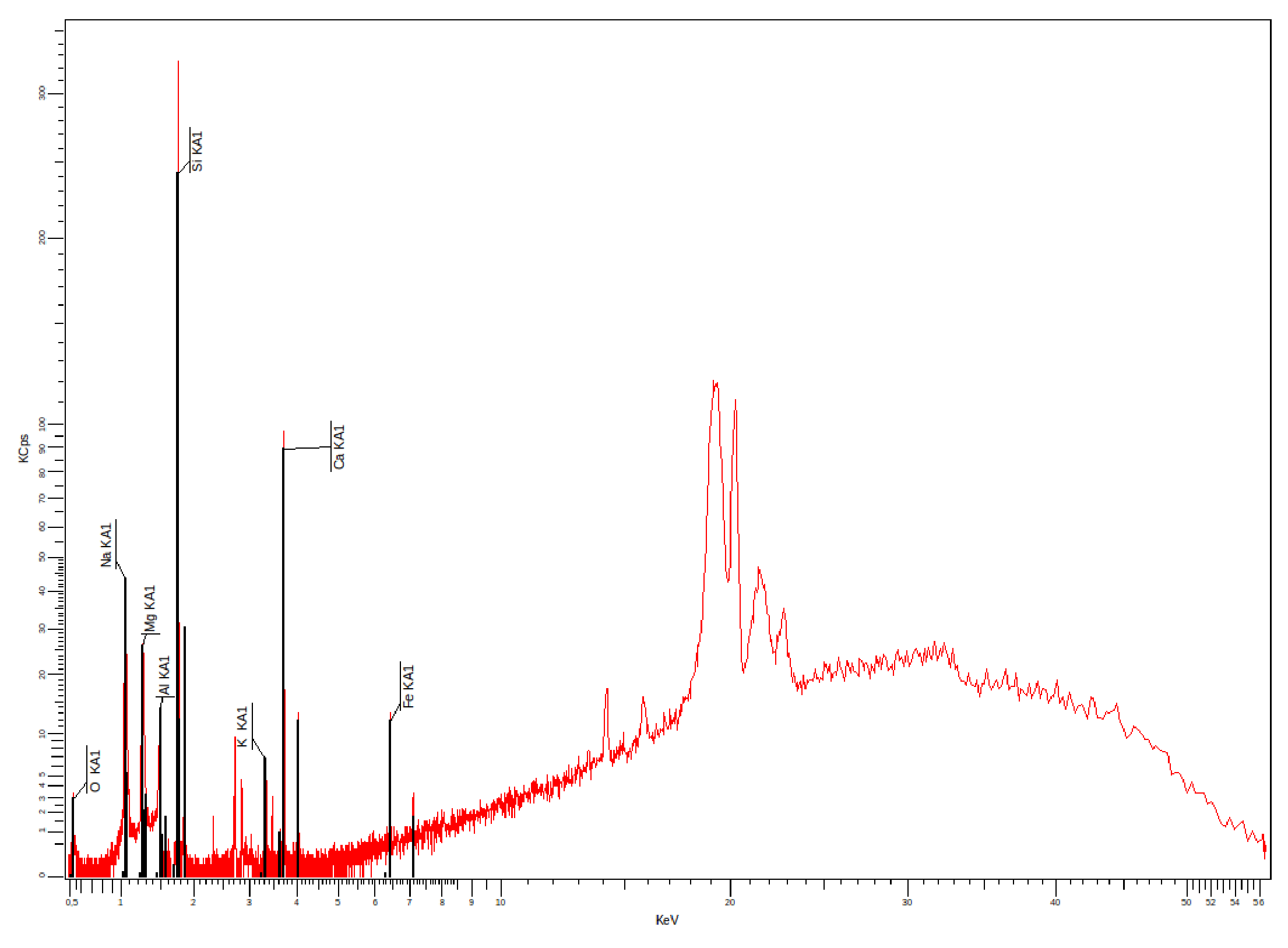}
\includegraphics[height=.16\textheight]{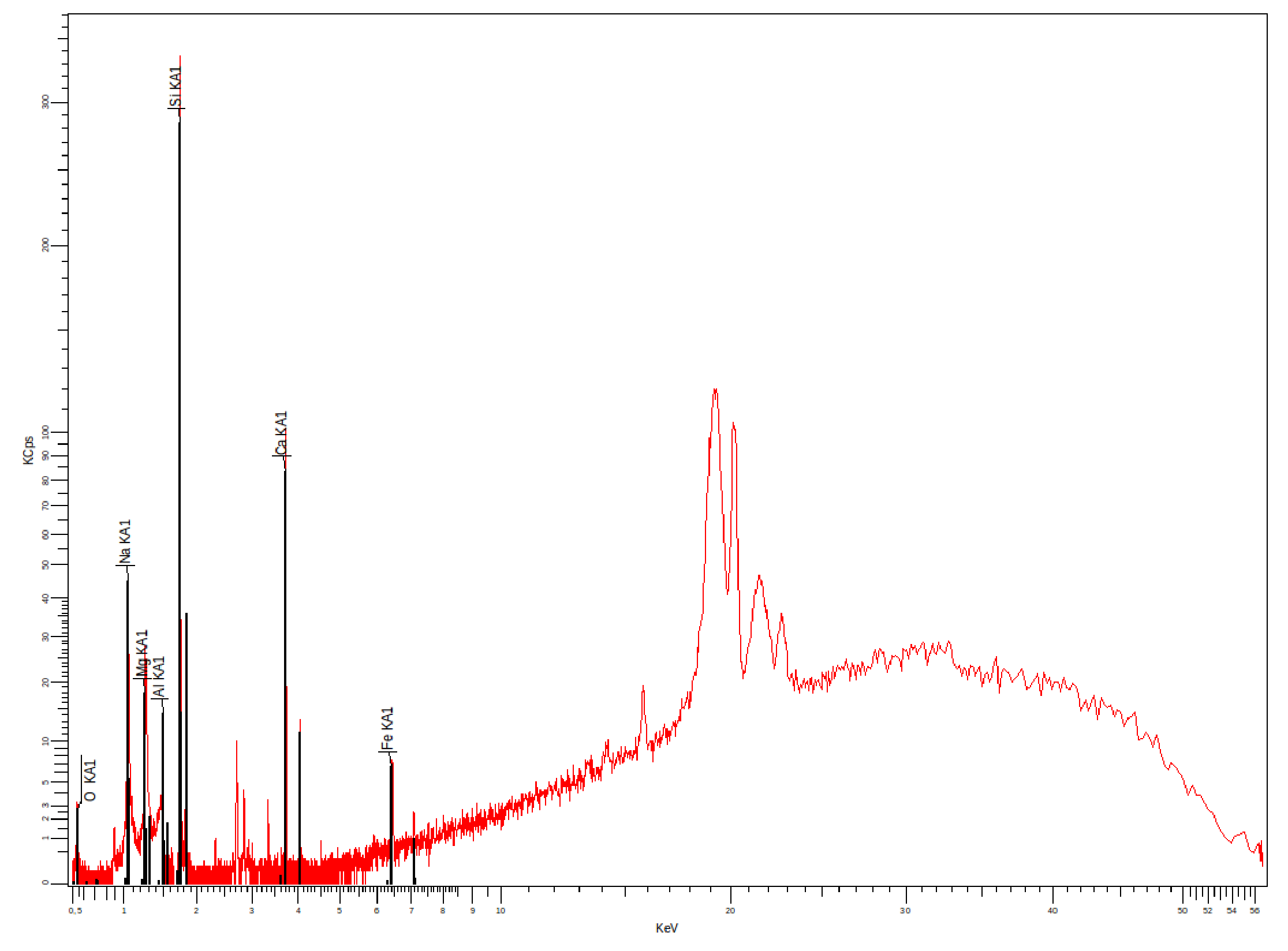}

  \caption{WD-XRF analysis of Asahi, Saint Gobain and Modi glass samples respectively. Note that x-axis denotes KeV and y-axis denotes KCps.}
\label{fig:WDXRF1}
\end{figure}

\begin{figure}[htbp]
\renewcommand{\figurename}{Fig.}
\centering
\includegraphics[height=.16\textheight]{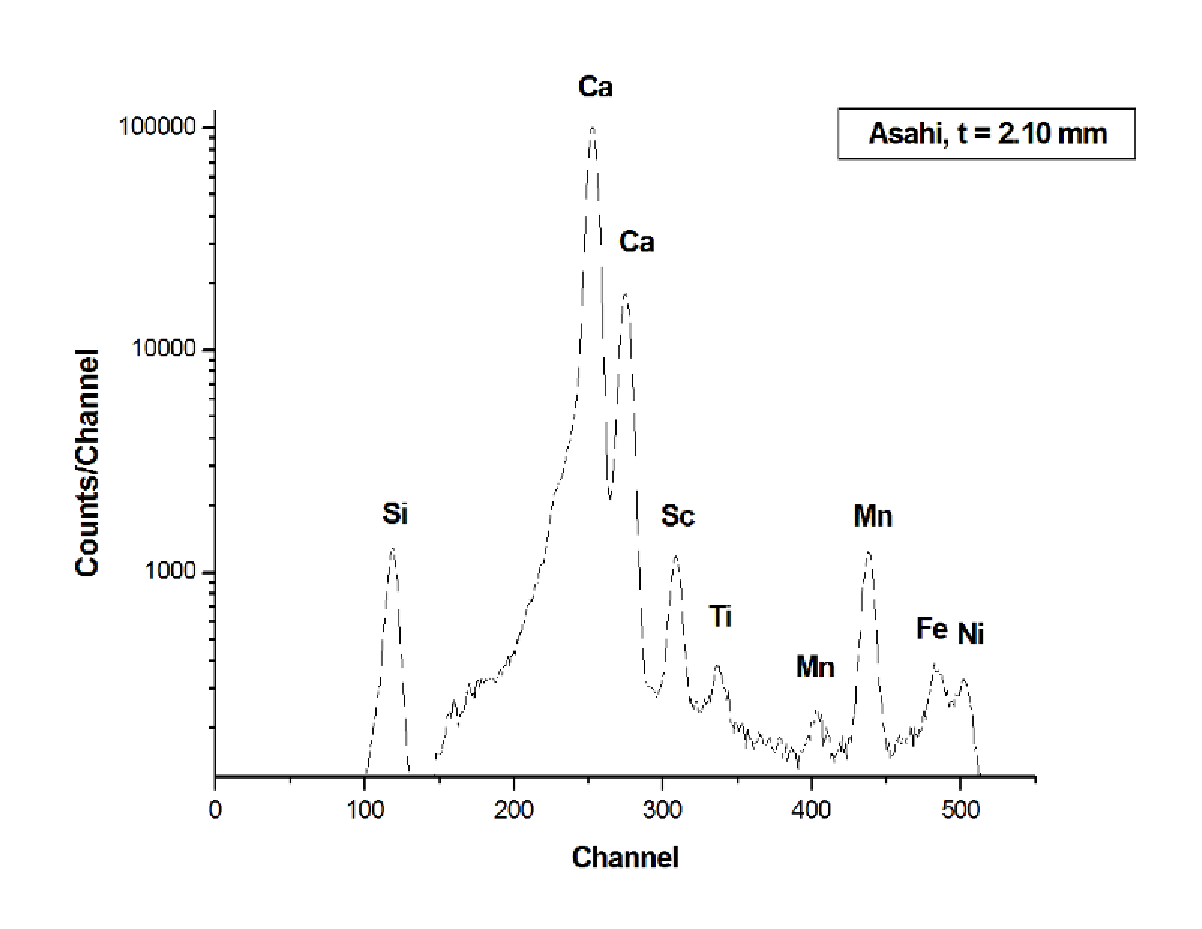}
\includegraphics[height=.16\textheight]{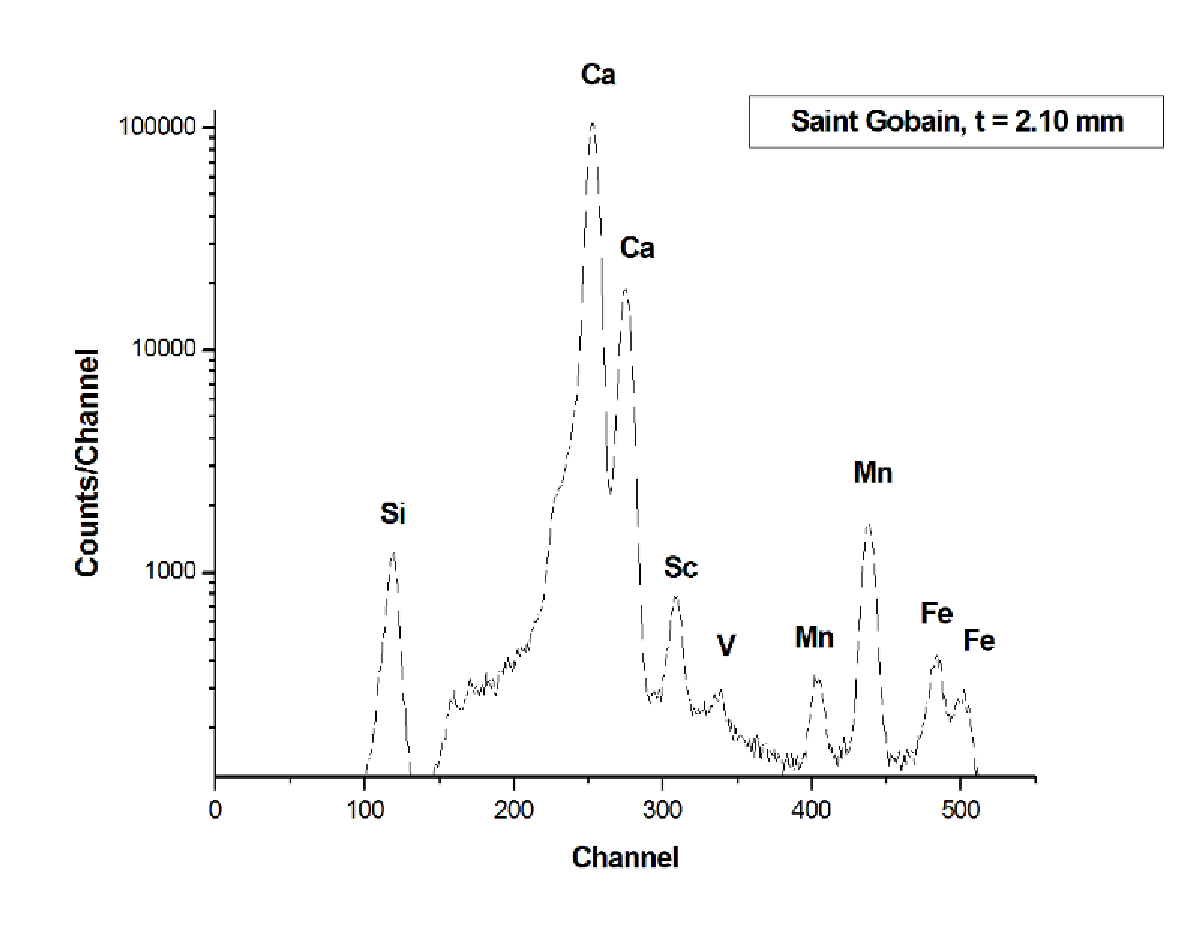}
 \includegraphics[height=.16\textheight]{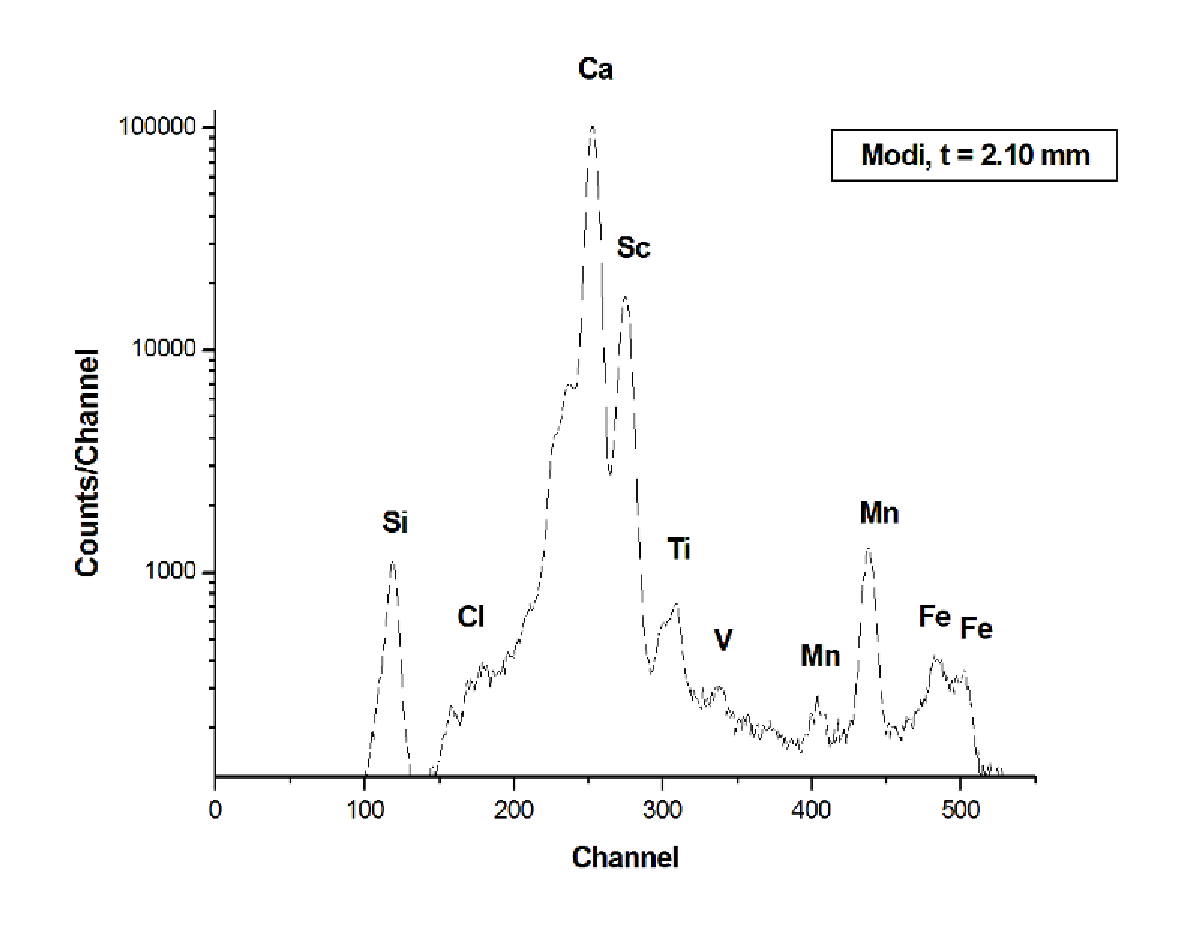}
  \caption{PIXE analysis of Asahi, Saint Gobain and Modi glass samples respectively.}
\label{fig:PIXE1}
\end{figure}

{\bf {Elemental and Compositional Studies}}: It is important to study the composition of the glass in order to get the information of elements or ions in the glass. The fractional percentages of weights of various compounds present in the glass samples were obtained using the Wavelength Dispersive X-ray Spectroscopy (WD-XRF) technique. PIXE (Proton Induced X-ray Emission spectroscopy) technique is used as supplementary to WD-XRF to do elemental analysis, which has been done using cyclotron. Fig.~\ref{fig:WDXRF1} shows the WD-XRF analysis and Fig.~\ref{fig:PIXE1} shows the PIXE analysis. Table~\ref{table2} shows the composition of all the samples. The percentages of $SiO_{2}$ and $Na_{2}O$ are important as they are the main compound of the glass and they show a constant percentage. $Sc_{2}O_{3}$ degrades the quality of Modi sample. 

Thus, it was concluded on the basis of optical, surface properties and elemental composition, that Asahi and Saint Gobain were better than Modi glass sample. The surface of Modi glass sample was observed to be of poor quality on the basis of SEM and AFM results and the Modi glass sample is impure than the Asahi and Saint Gobain. This will effect the performance of RPC made of Modi glass sample. Now we describe the fabrication and characterisation of the RPCs made of these glass samples.

\begin{table}[ht]
\centering 
\scalebox{0.8}{
\begin{tabular}{|c c c c|} 
\hline 
Compound & Modi($\%$) & SG($\%$) & Asahi($\%$)\\
\hline                  
$SiO_{2}$ & 72.43 & 73.03 & 71.84 \\ 
$Na_{2}O$ & 13.04 & 13.05 & 14.04 \\
$CaO$ & 9.13 & 8.45 & 8.96 \\
$MgO$ & 3.85 & 3.72 & 4.06  \\
$Al_{2}O_{3}$ & 0.94 & 0.79 & 0.69 \\
$SO_{3}$ & 0.13 & 0.23 & 0.15 \\
$Fe_{2}O_{3}$ & 0.10 & 0.18 & 0.11 \\
$K_{2}O$ & 0.25 & 0.48 & $-$ \\
$TiO_{2}$ & 0.06 & $-$ & 0.09 \\
$ZrO_{2}$ & 99 PPM & 40 PPM & 0.02 \\
$SrO$ & $-$ & 0.01 & $-$ \\
$Sc_{2}O_{3}$ & 0.02 & $-$ & $-$ \\
$Cl$ & $-$ & 0.03 & 0.03 \\
\hline
\end{tabular}
}
\caption{Composition of glass samples.} 
\label{table2} 
\end{table}


\section{Fabrication and Characterisation of glass RPCs}
\label{virpc}

After characterising the glass samples of different manufacturers, all the different glass samples were used to fabricate RPC to measure efficiency as a second order cross check. We describe the process for fabricating one such RPC. 

The characterisation of various glass samples of the different manufacturers procured from domestic market further motivated us to check their performance by characterisation and efficiency measurements of the RPCs made of the tested glass samples. Therefore, we fabricated RPCs of Asahi, Saint Gobain and Modi glass of 2.10 mm thickness \cite{Bhide, Datar} to check their performance. One of the RPC parameters called ``strip width'' study is important in order to do precise physics analysis. Depending on the physics goals, the strip width of read-out boards (pick-up panels) can be optimised. For this, we have studied the efficiency and cross-talk of the RPCs and varied strip width to check their performance \cite{strip_kan}.

\begin{figure}[ht]
\renewcommand{\figurename}{Fig.}
\centering
\includegraphics[height=.19\textheight]{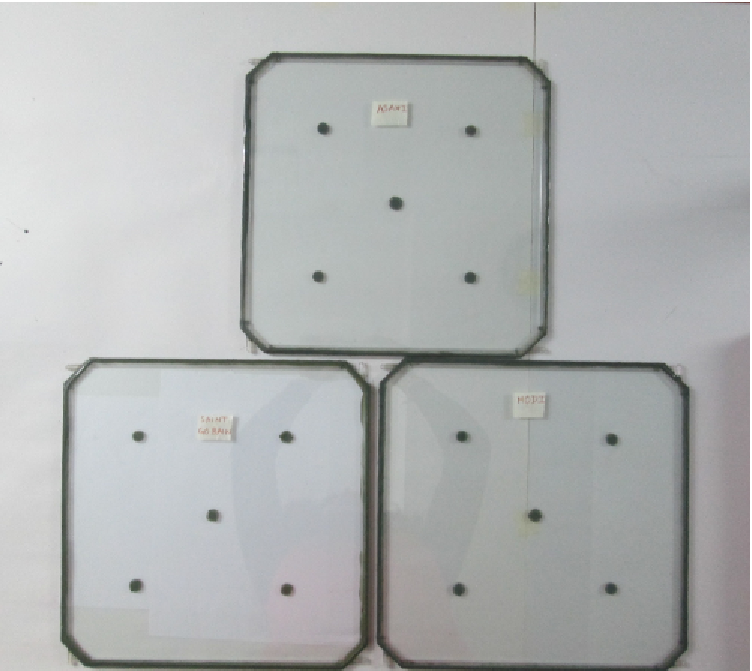}
 \includegraphics[height=.19\textheight]{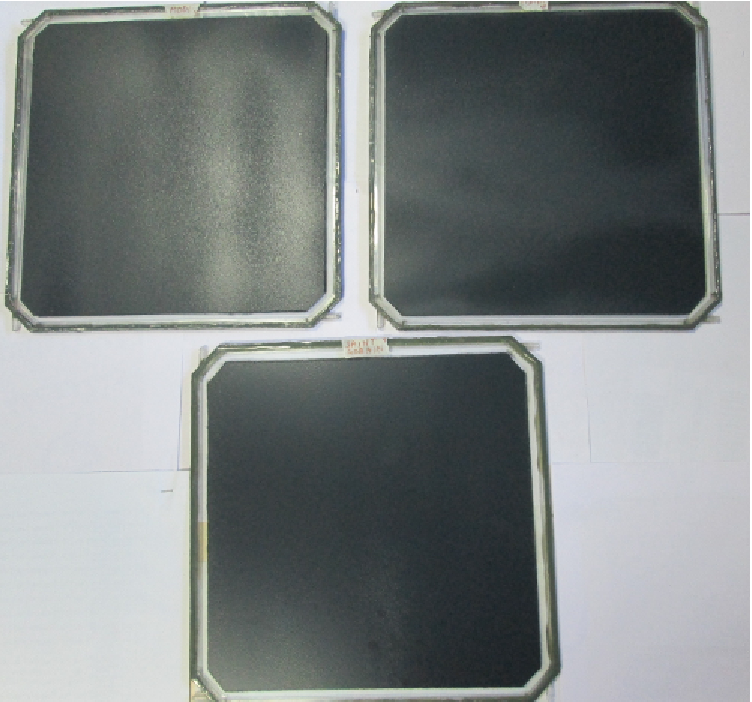}
  \caption{The fabricated RPCs, before graphite coating (left) and after graphite coating (right).}
\label{fig:SG-Asahi-Modi}
\end{figure}

We have fabricated RPCs of Asahi, Saint Gobain and Modi glass using standardized procedure which is as follows. The two glass plates of 30 cm $\times$ 30 cm size with the four corners chamfered at $45^{0}$ were cleaned with distilled water and ethanol. A drop of glue (DP 125 grey) was applied on one side of the surface at four equidistant positions and one at the center. The button spacers were placed on the top of each glue drop and pressed then. All the side spacers were placed covering all the sides of the glass and nozzles were kept at the corners in a manner such that their direction is in either clockwise direction or anti-clockwise direction. All the side spacers and nozzles were glued with one glass plate and waited for almost 12 hours to dry them at room temperature. After drying up, a drop of glue was applied on each button spacers and placed the second glass gently on the top of the button spacers. Glue was applied again on all the sides between side spacers, nozzles and glass plates. A weight was put on the RPC and allowed the glue to harden for more than 12 hours. Then the unpainted glass RPC was painted manually with spray gun on the outer surface with graphite paint. The painted RPC was allowed to dry at room temperature. Fig.~\ref{fig:SG-Asahi-Modi} shows the fabricated RPCs (Asahi, Saint Gobain and Modi) before and after graphite coating. 

\begin{figure}[htbp]
\renewcommand{\figurename}{Fig.}
\centering
\includegraphics[height=.18\textheight]{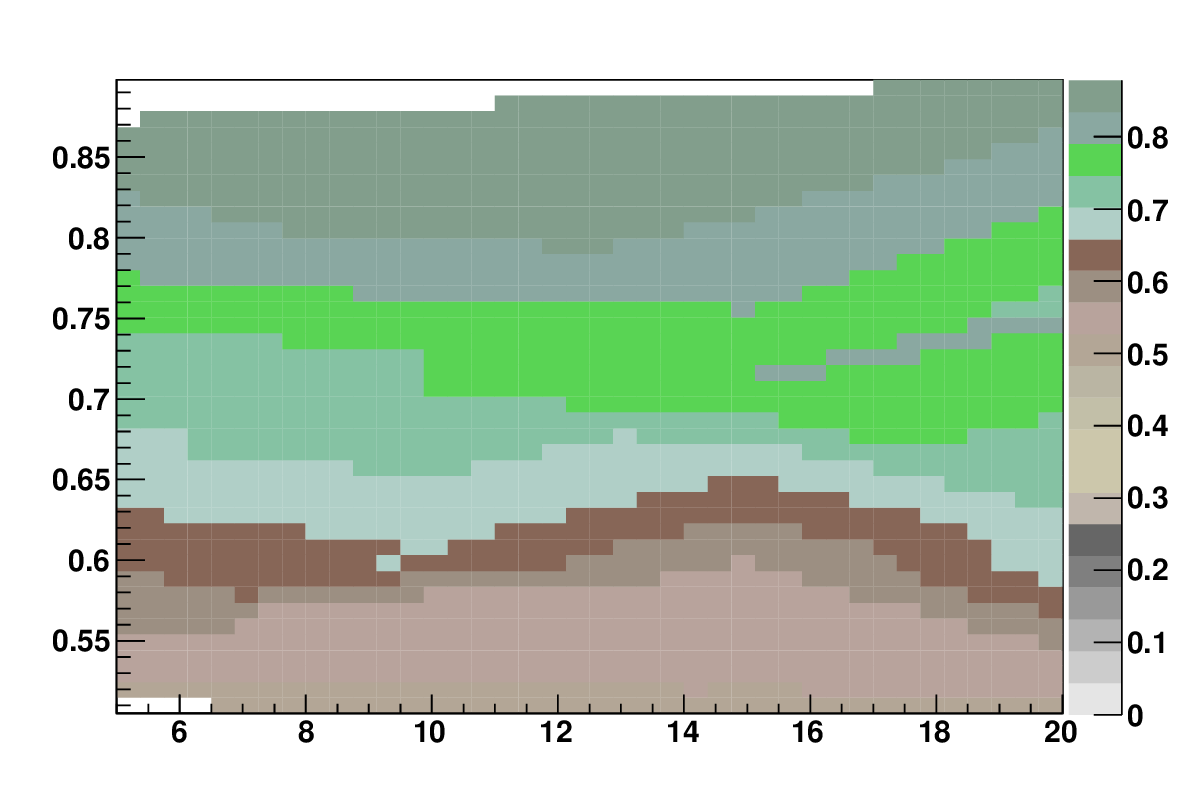}
 \includegraphics[height=.18\textheight]{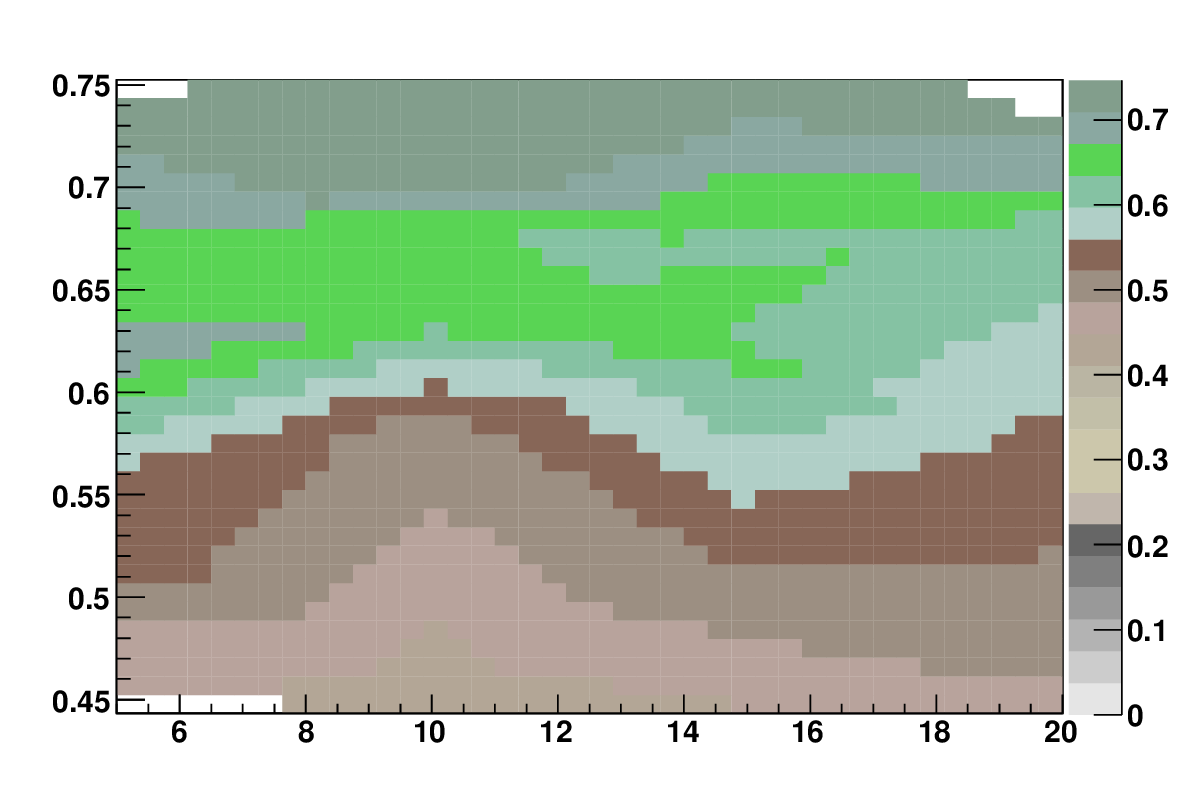}
  \caption{Surface resistivity of Asahi (left) and Modi (right) glass RPC.}
\label{fig:SR}
\end{figure}

We measured the surface resistivity of the fabricated RPCs which was of the order of around 600--800 k$\Omega$ for Asahi, Saint Gobain glass RPC and 600--700 k$\Omega$ for Modi glass RPC as shown in Fig.~\ref{fig:SR}. The best and the worst resistivity of glasses is shown in the figure. The variation in the bulk resistivity is of order of 10--15\%. However, the variation in the surface resistivity is due to the manual spray painting. We fabricated pick-up panels also of different strip size: 1.8 cm, 2.8 cm and 3.8 cm for doing our strip width studies \cite{strip_kan}. The pick-up panel is made from plastic honeycomb of 5 mm thick with 50 micron aluminium sheet (for grounding) on one side and copper strips of 2.8 cm (or 1.8 or 3.8 cm) with a gap of 0.2 cm on the other side. Each strip is terminated with a 50 $\Omega$ impedance to match the characteristic impedance of the preamplifier. Other end of the strip is soldered to wire to connect with electronics. Gas leakage and pressure test on these RPCs were done using standard techniques \cite{kalmani:2009}. These RPCs were characterised for V-I, efficiency and cross-talk which is described in the next subsection. 

\subsection{V-I Characteristics of RPC}

\begin{figure}[htbp]
\renewcommand{\figurename}{Fig.}
\centering
\includegraphics[height=.2\textheight]{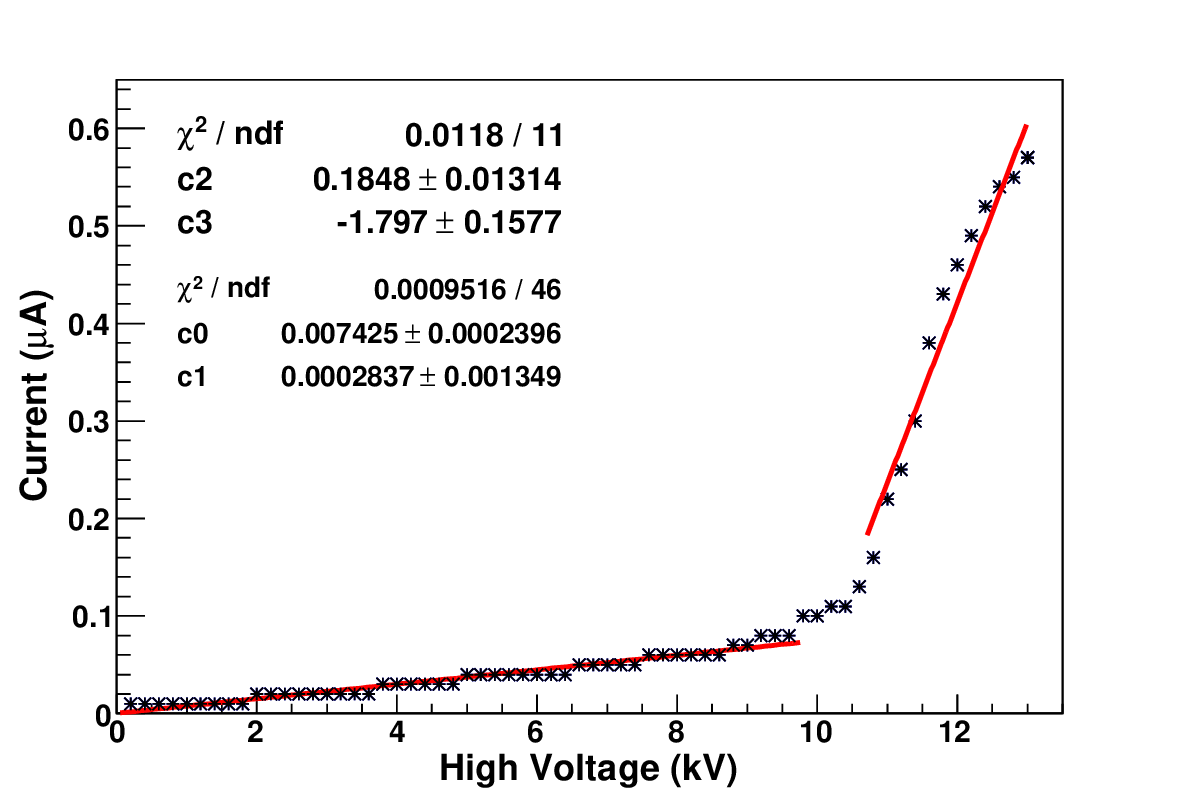}
 \includegraphics[height=.2\textheight]{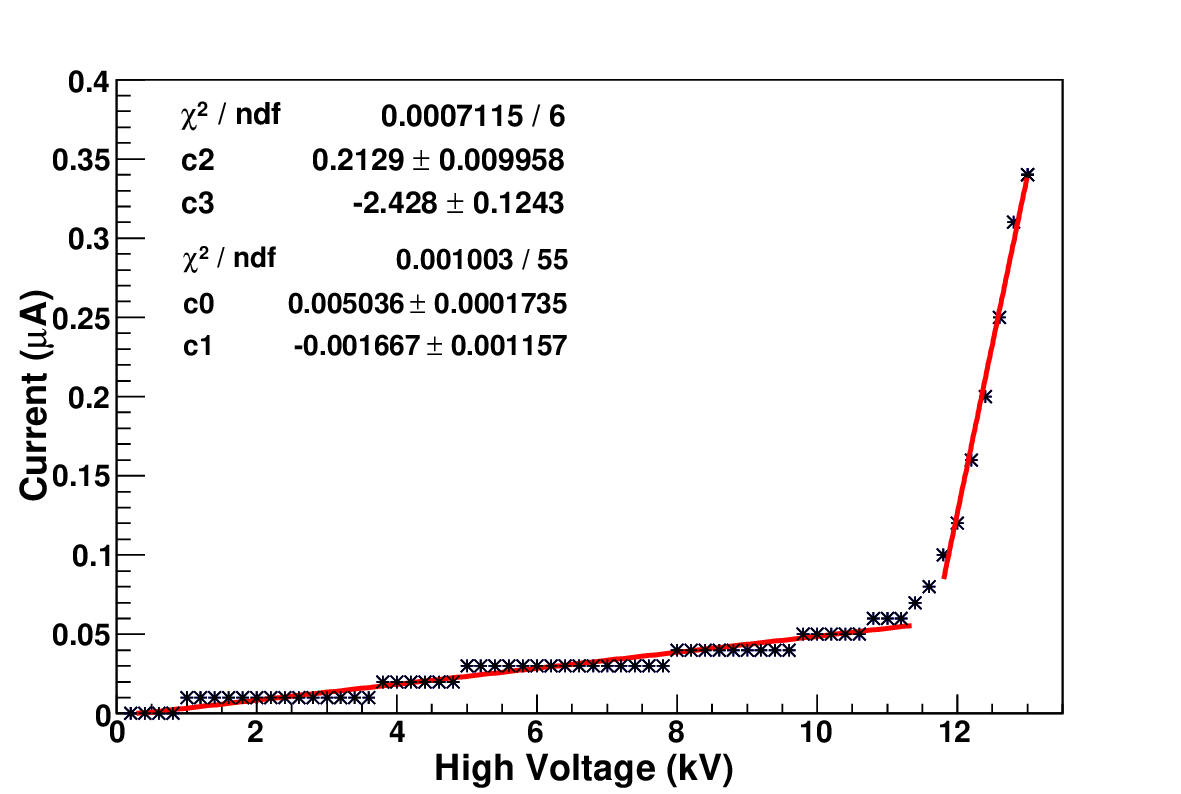}
\includegraphics[height=.2\textheight]{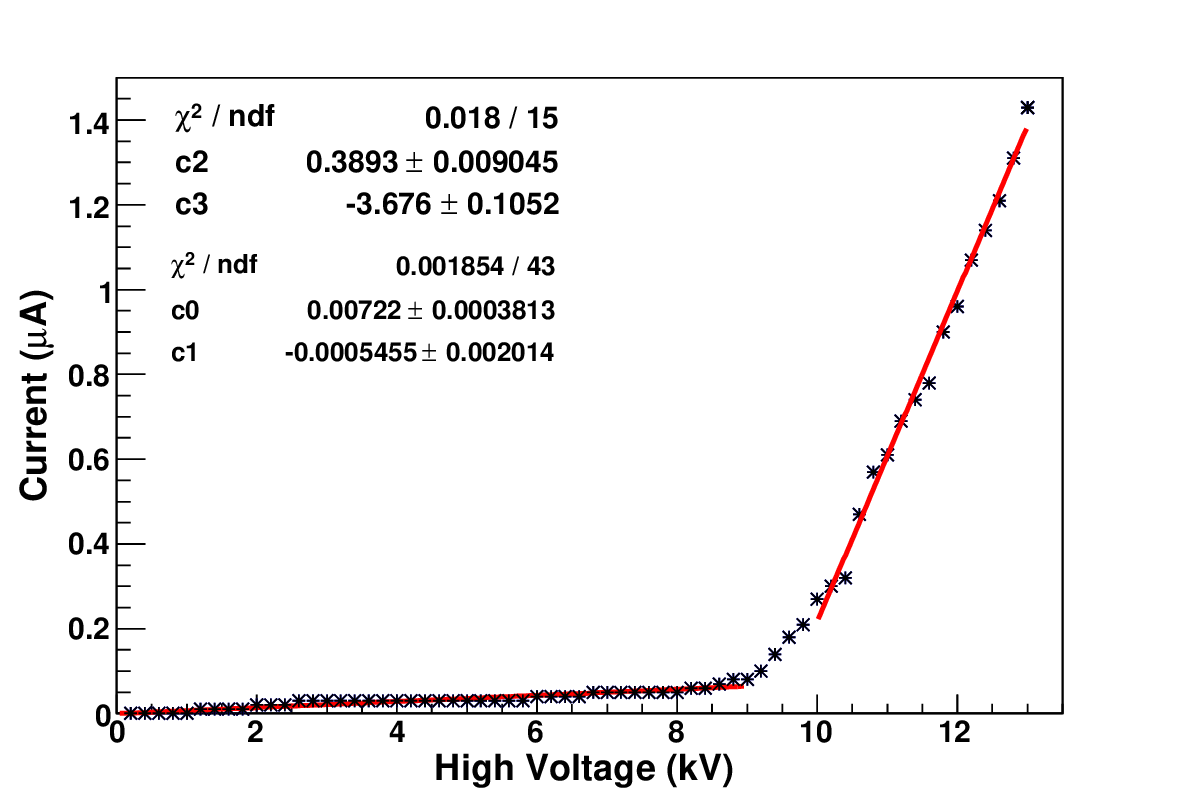}
  \caption{V-I characteristics of Asahi glass RPC with two gases, three gases in avalanche mode and streamer mode respectively.}
\label{fig:VI-Asahi}
\end{figure}

\begin{figure}[htbp]
\renewcommand{\figurename}{Fig.}
\centering
\includegraphics[height=.2\textheight]{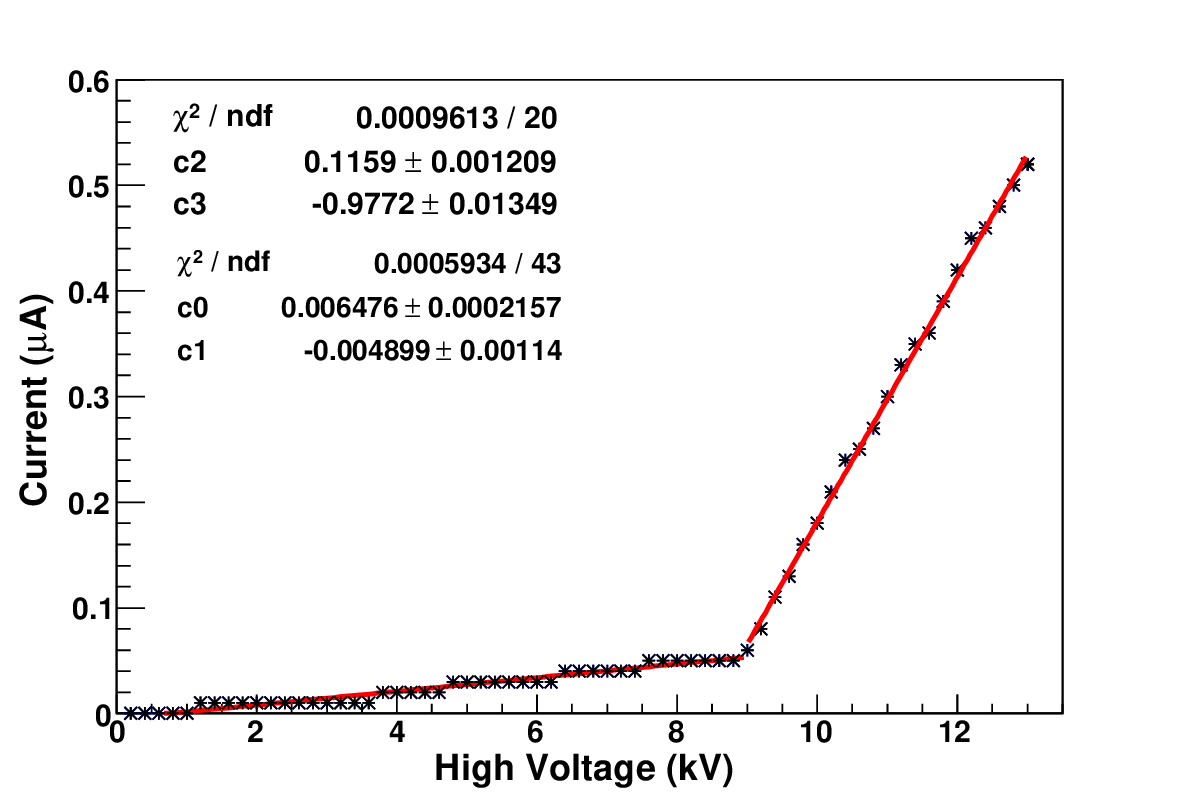}
 \includegraphics[height=.2\textheight]{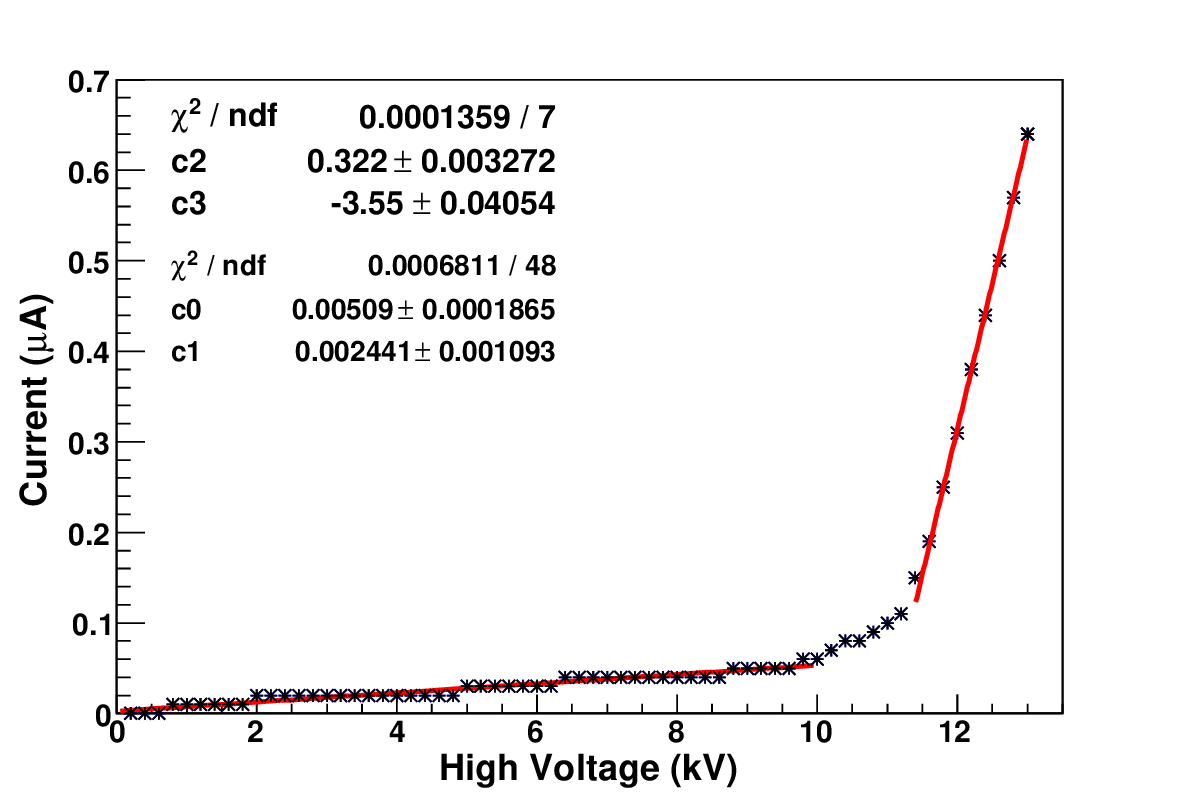}
\includegraphics[height=.2\textheight]{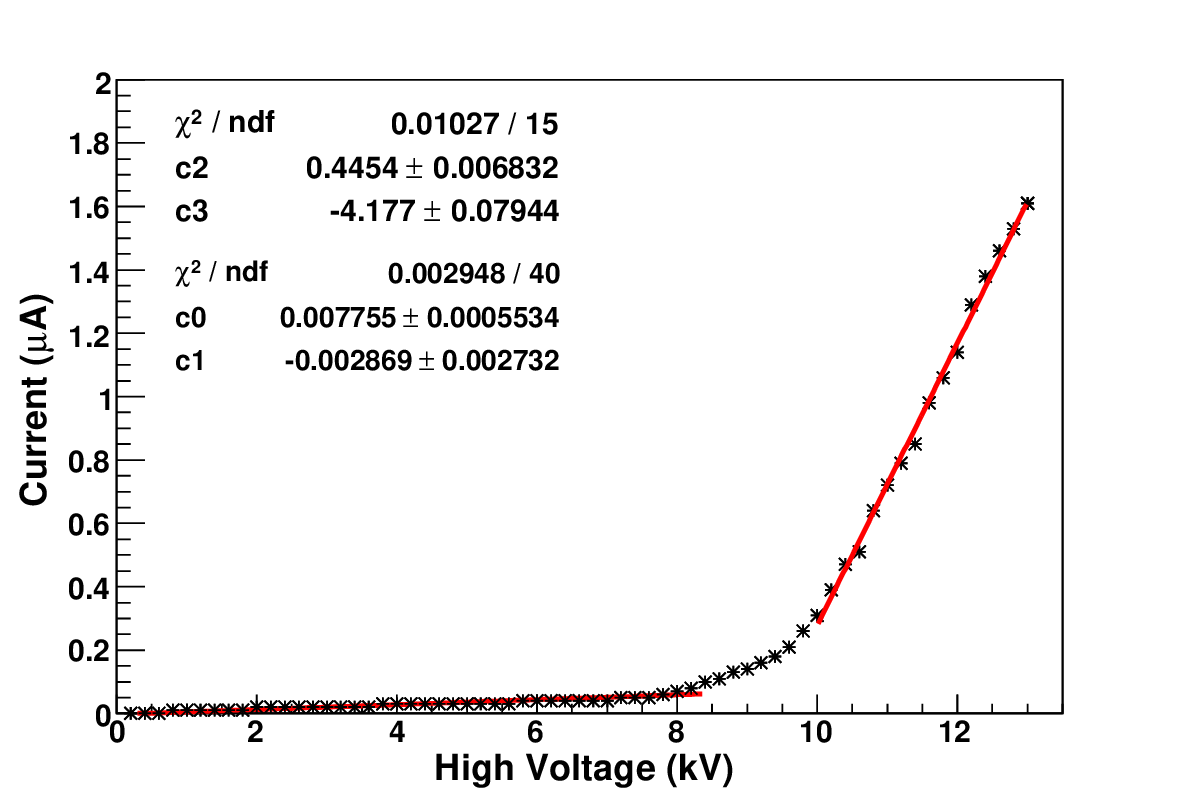}
  \caption{V-I characteristics of Saint Gobain glass RPC with two gases, three gases in avalanche mode and streamer mode respectively.}
\label{fig:VI-SG}
\end{figure}

\begin{figure}[htbp]
\renewcommand{\figurename}{Fig.}
\centering
\includegraphics[height=.2\textheight]{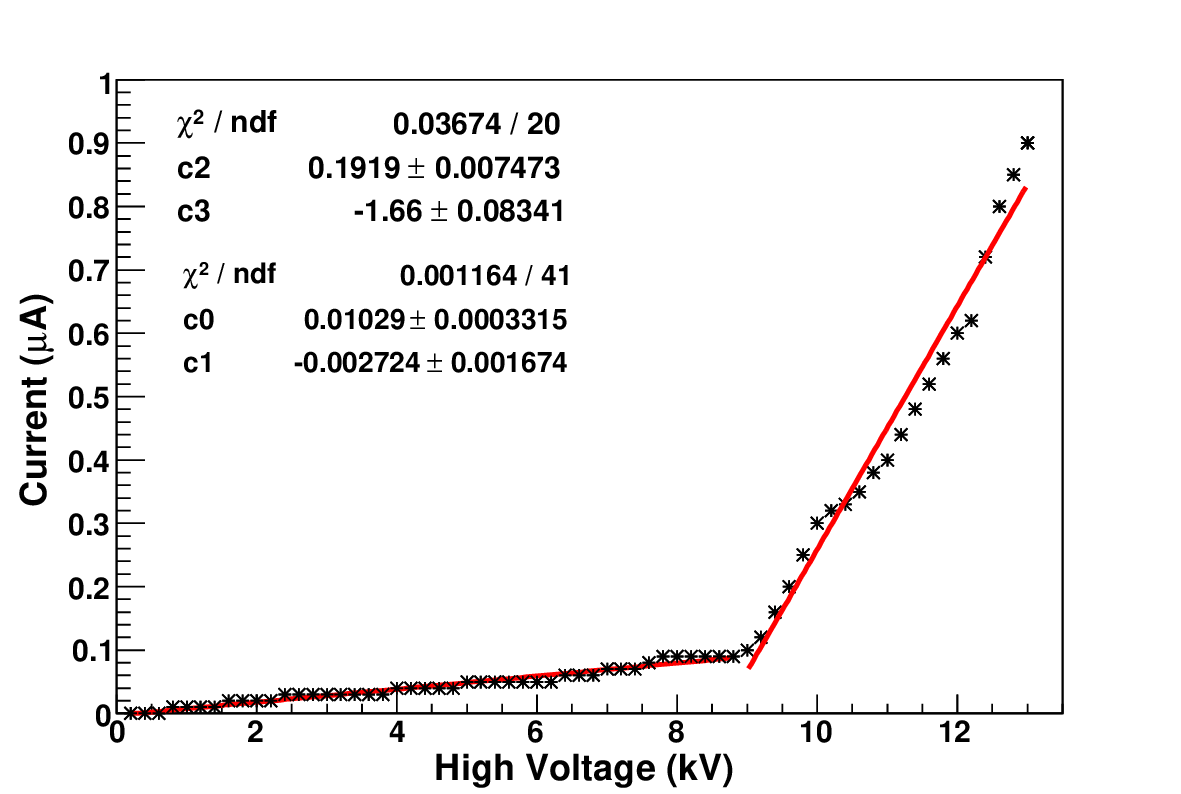}
 \includegraphics[height=.2\textheight]{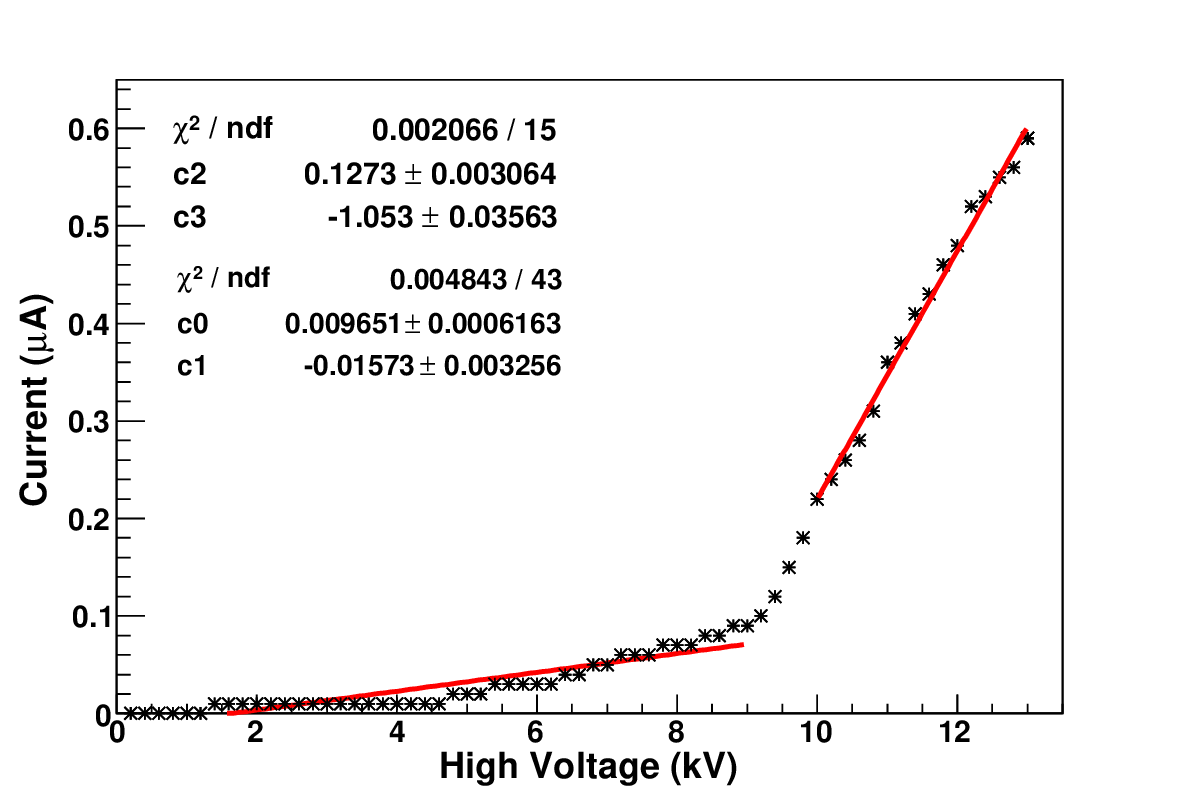}
\includegraphics[height=.2\textheight]{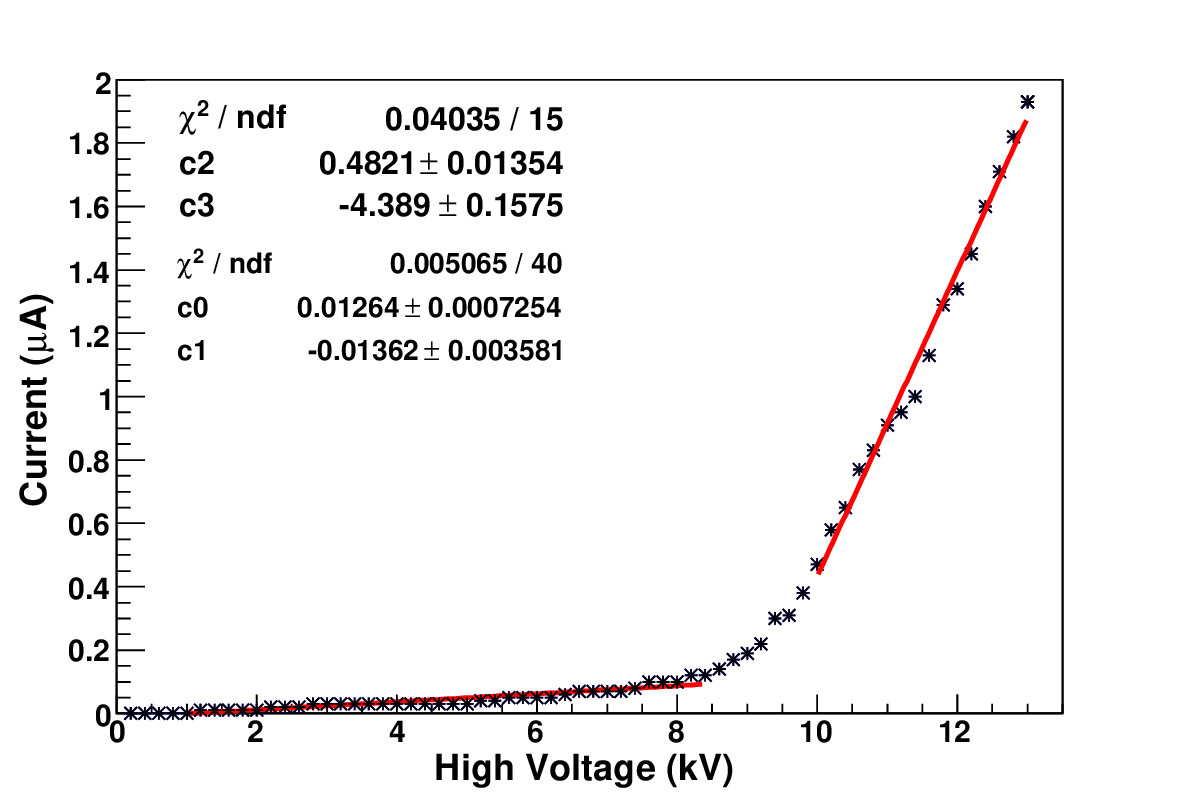}
  \caption{V-I characteristics of Modi glass RPC with two gases, three gases in avalanche mode and streamer mode respectively.}
\label{fig:VI-Modi}
\end{figure}

The packed RPCs were characterised for the leakage current with different modes. The two modes that have been used are: avalanche and streamer for three glass RPCs: Asahi, Saint Gobain and Modi. In avalanche mode, we tested firstly the RPCs with two gases in the ratio \cite{Mengucci} --~\cite{abe} Freon(R134a) : Isobutane :: 95.5 : 4.5 and obtained their V-I. Later, we added a quenching gas $SF_{6}$ in order to check the performance of the RPCs. The gas composition used for this was Freon(R134a) : Isobutane : $SF_{6}$ :: 95.15 : 4.51 : 0.34. We also characterised RPCs in streamer mode with the gas ratio taken as Freon(R134a) : Isobutane : Argon :: 62 : 8 : 30. We have categorized the figures of V-I into three sets for Asahi, Saint Gobain and Modi glass RPC with two gases, three gases (avalanche mode) and streamer mode respectively as shown in Figs.~\ref{fig:VI-Asahi}, ~\ref{fig:VI-SG}, ~\ref{fig:VI-Modi}.

In the V-I plot, both the ohmic and non-ohmic regions are clearly seen. At the lower voltage (LV), the primary ionization does not produce avalanche. So, the gas gap impedance is infinite, therefore the current through the RPC is proportional to resistance provided by spacer which is less than the gap resistance. While at higher voltages (HV), when avalanches are produced the gas resistance drops down and the current obtained is due to glass plates. Table~\ref{table3} shows the resistances on the basis of the different voltage regions obtained from V-I plots. Resistance is low in the lower voltage region as the resistance is due to spacers but its high in the higher voltage region as the voltage is due to glass plates. In the next subsection we describe the efficiency and cross-talk measurements taken for the various RPCs.

\begin{table}[ht]
\centering  
\scalebox{0.8}{
\begin{tabular}{|c c c c|} 
\hline 
RPC & Asahi($M\Omega$) & Saint Gobain($M\Omega$) & Modi($M\Omega$)\\ 
\hline                 
2 gases
(Avalanche) (LV) & 5.41 & 8.62 & 5.21 \\
2 gases (Avalanche) (HV) & 134.68 & 154.41 & 97.18 \\
3 gases (Avalanche) (LV) & 4.69 & 3.10 & 7.85 \\
3 gases (Avalanche) (HV) & 198.57 & 196.46 & 103.61 \\
Streamer (LV) & 2.56 & 2.24 & 2.07 \\
Streamer (HV) & 138.50 & 128.94 & 79.11 \\
\hline
\end{tabular}
}
\caption{Resistances on the basis of the different voltage regions.}
\label{table3} 
\end{table}

\subsection{Cross-talk and Efficiency of Various Glass RPCs}

For performing the cross-talk measurements we used cosmic ray muon test stand. The muon ionizes the gas in RPC and the signal generated is picked up by copper strips of the pick-up panels. To transfer this signal to the electronics, we need preamplifiers in case of avalanche mode only. We used charge-sensitive fast preamplifier with the gain of 75 $\%$. The pre-amplified RPC signals are fed to AFE (analog front-end) boards in order to convert the amplified analog RPC pulses into logic signals by using a low threshold discriminator circuits. The discriminator signals from the AFE boards are further processed by a DFE (digital front-end) for multiplexing of these signals and the actual counting was done by scalars. When a trigger signal from scintillator detector was received, the processed signals were latched and recorded by the NIM and CAMAC based back-end electronics: control and readout module. These modules were interfaced to a PC through CAMAC controller which regulated the synchronous functioning of all CAMAC modules.

\begin{figure}[htbp]
\renewcommand{\figurename}{Fig.}
\centering
\includegraphics[height=.35\textheight]{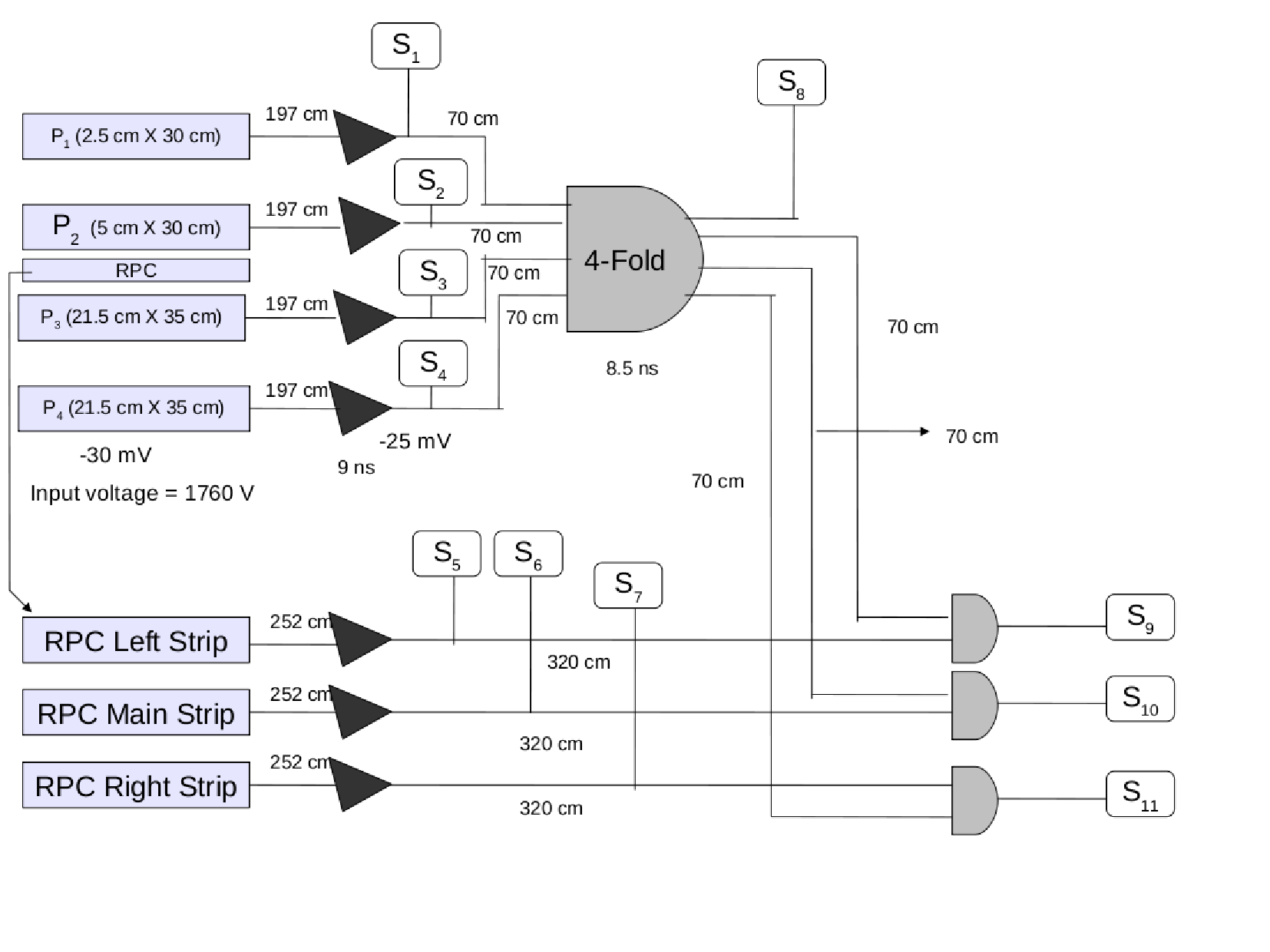}
  \caption{Trigger scheme circuit diagram for the testing of RPCs.}
\label{fig:trigger-RPC}
\end{figure}

\noindent {\bf Efficiency}
Fig.~\ref{fig:trigger-RPC} shows trigger scheme circuit diagram for testing RPC. The discriminators are connected to the paddles $P_{1}$, $P_{2}$, $P_{3}$ and $P_{4}$ to make 4-fold coincidence (to form a trigger signal). This 4-fold along with the different RPC strips form a coincidence. $P_{1}$ = 2.5 cm $\times$ 30 cm, $P_{2}$ = 5 cm $\times$ 35 cm, $P_{3}$ = 21.5 cm $\times$ 35 cm, $P_{4}$ = 21.5 cm $\times$ 35 cm. $P_{1}$ and $P_{2}$ were placed one above the other in a manner to create a window of about 14 cm $\times$ 2.5 cm. They were aligned with strip number 3 which was labelled as ``main strip'', whereas $2^{nd}$ strip was labelled as ``left strip'' and $4^{th}$ strip was labelled as ``right strip''. The data was obtained for X-strip of RPC. The pulse width of scintillators were kept at 60 ns and RPCs at 50 ns. The RPC with strip width 2.8 cm of pick-up panel was placed after $P_{2}$ to ensure that when the muon passed through all the four paddles it passed through the RPC too. The counters, $S_{1}$ to $S_{7}$ are connected to count muon events that passed through scintillator detectors. One preamplifier board with 8 capacity was used to read the pick-up strips. Temperature at $24.5^\circ$C and relative humidity $RH = 40\%$ was maintained. But the source of error was the opening and closing of the door which caused moisture level sometimes to go up and affected the efficiency of the RPCs.

All the paddles were ANDed ($P_{1}$.$P_{2}$.$P_{3}$.$P_{4}$) to give a 4-fold signal, which acted as the trigger pulse. We assumed the passage of muon through the set window, if this ANDed signal was one. The trigger signal and main strip signal from the RPC was then ANDed together. When the trigger and the respective RPC strip signal was one then only the scalar counter of respective strip incremented its count by one. The number of time this condition was satisfied gave efficiency and is given by,

\begin{equation}
\hbox{Efficiency} = \frac{( \hbox{No. of strip hits and Trigger})}
		{\hbox{Total  no. trigger}}~.
\label{eff_rpc}
\end{equation}

\noindent {\bf Cross-talk}
The fluctuations in the efficiency were due to the ``noise rate''. It is defined as the rate at which random noise signal hits the RPC strip. It can be due to cosmic ray particles, stray radioactivity and dark current in the chamber. Expecting that the RPC strip was aligned with window of cosmic ray telescope to pick up the signal. If it is picked up by adjacent strips, it is known as ``cross-talk'' and it can be due to misalignment of strip or due to inadequate amount of quenching gas used. An effort was put to reduce it to improve the efficiency and hence performance of the RPC.

\begin{figure}[htbp]
\renewcommand{\figurename}{Fig.}
\centering
\includegraphics[height=.2\textheight]{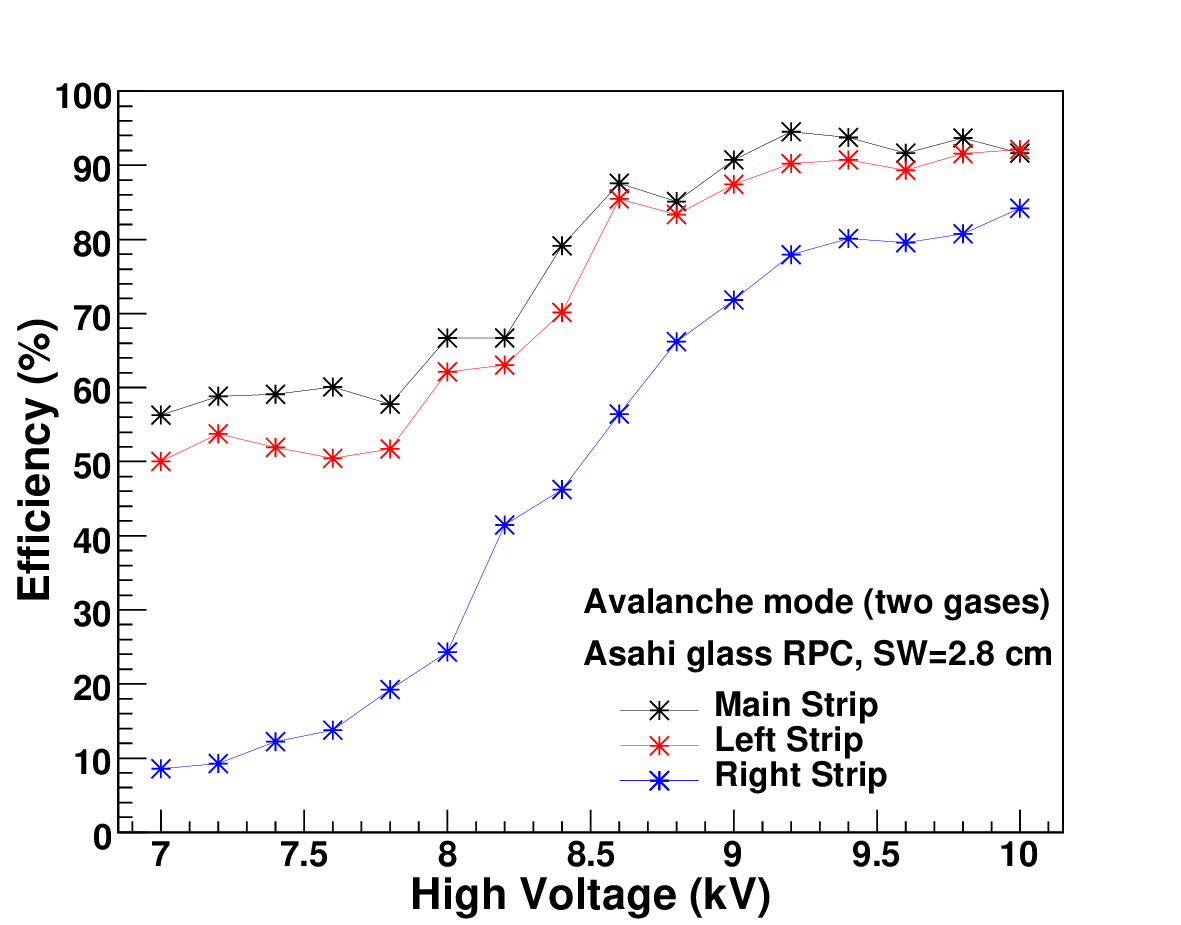}
 \includegraphics[height=.2\textheight]{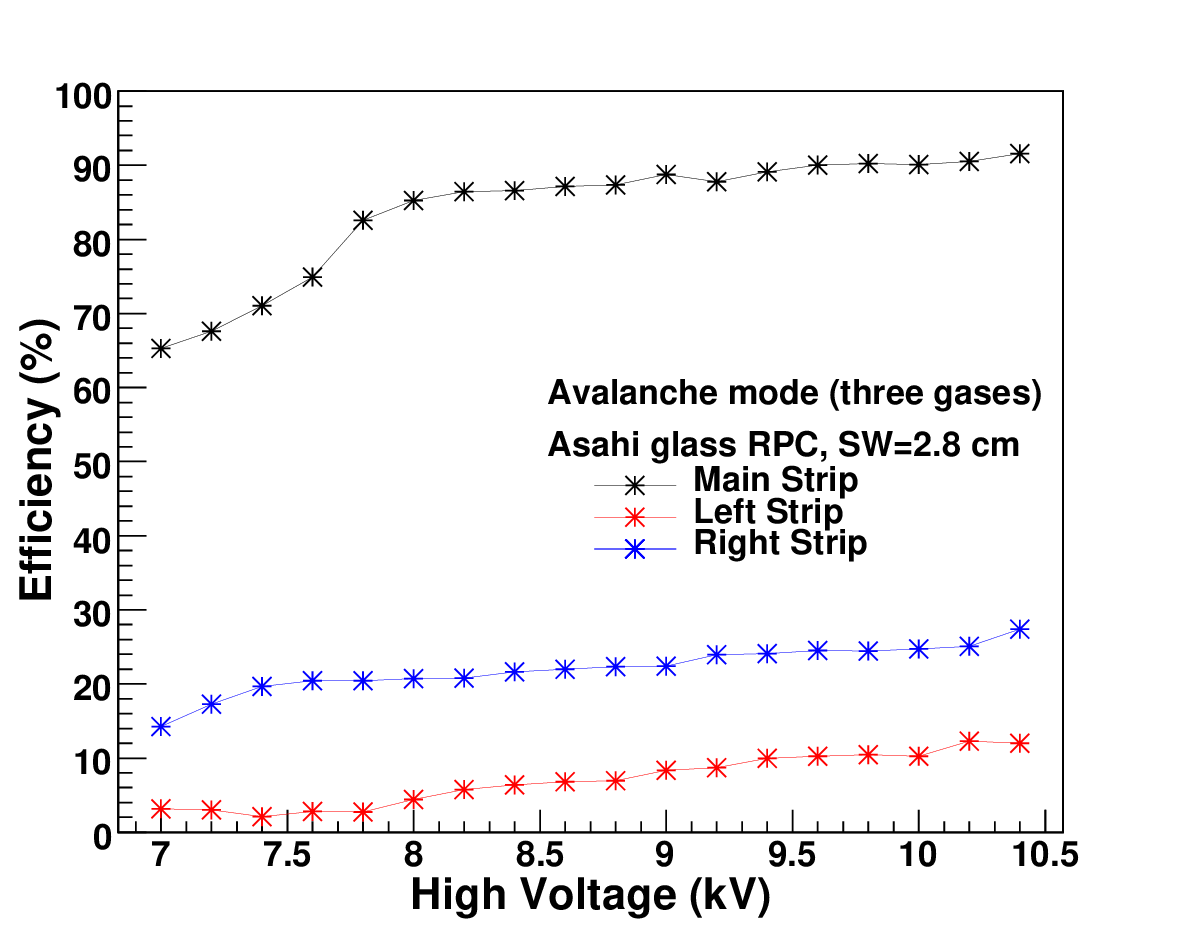}
\includegraphics[height=.2\textheight]{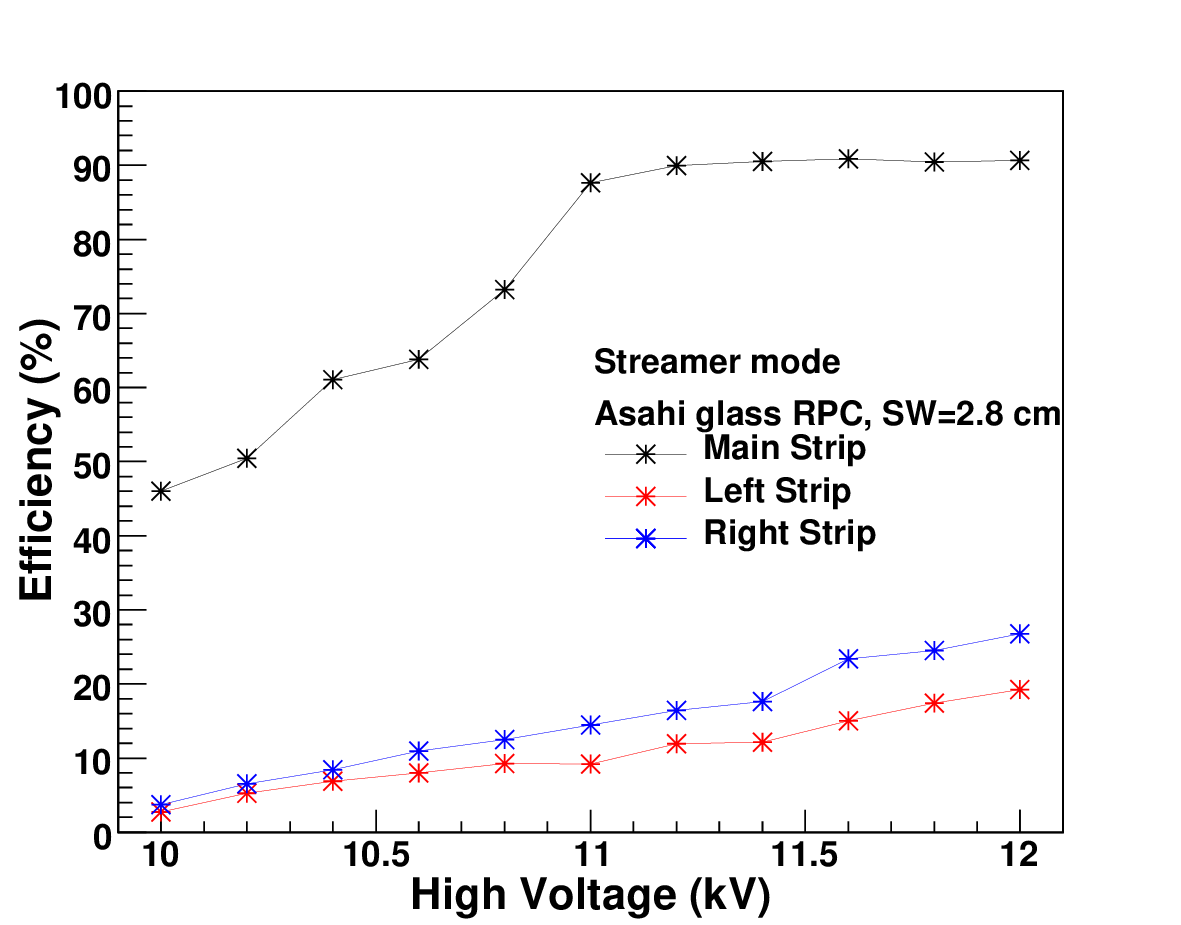}
  \caption{Cross-talk and Efficiency of Asahi glass RPC with the strip width of 2.8 cm in avalanche mode (two gases), avalanche mode (three gases) and streamer mode respectively.}
\label{fig:Asahi-2.8cm}
\end{figure}

Using the concept discussed above we obtained the efficiency and cross-talks of Asahi, Saint Gobain and Modi glass RPC for strip width of 2.8 cm. Fig.~\ref{fig:Asahi-2.8cm} show the efficiency and cross-talk of Asahi glass RPC with the strip width of 2.8 cm in avalanche mode (two gases), avalanche mode (three gases) and streamer mode respectively. From the figures it is observed that the efficiency of the detector increases with voltage reaching a plateau at higher voltage with efficiency greater than 90\%. The fluctuations may be due to the noise rate, especially for the two gases. The fluctuations in the efficiencies reduces with the addition of third gas in both avalanche mode and streamer mode.

\begin{figure}[htbp]
\renewcommand{\figurename}{Fig.}
\centering
\includegraphics[height=.2\textheight]{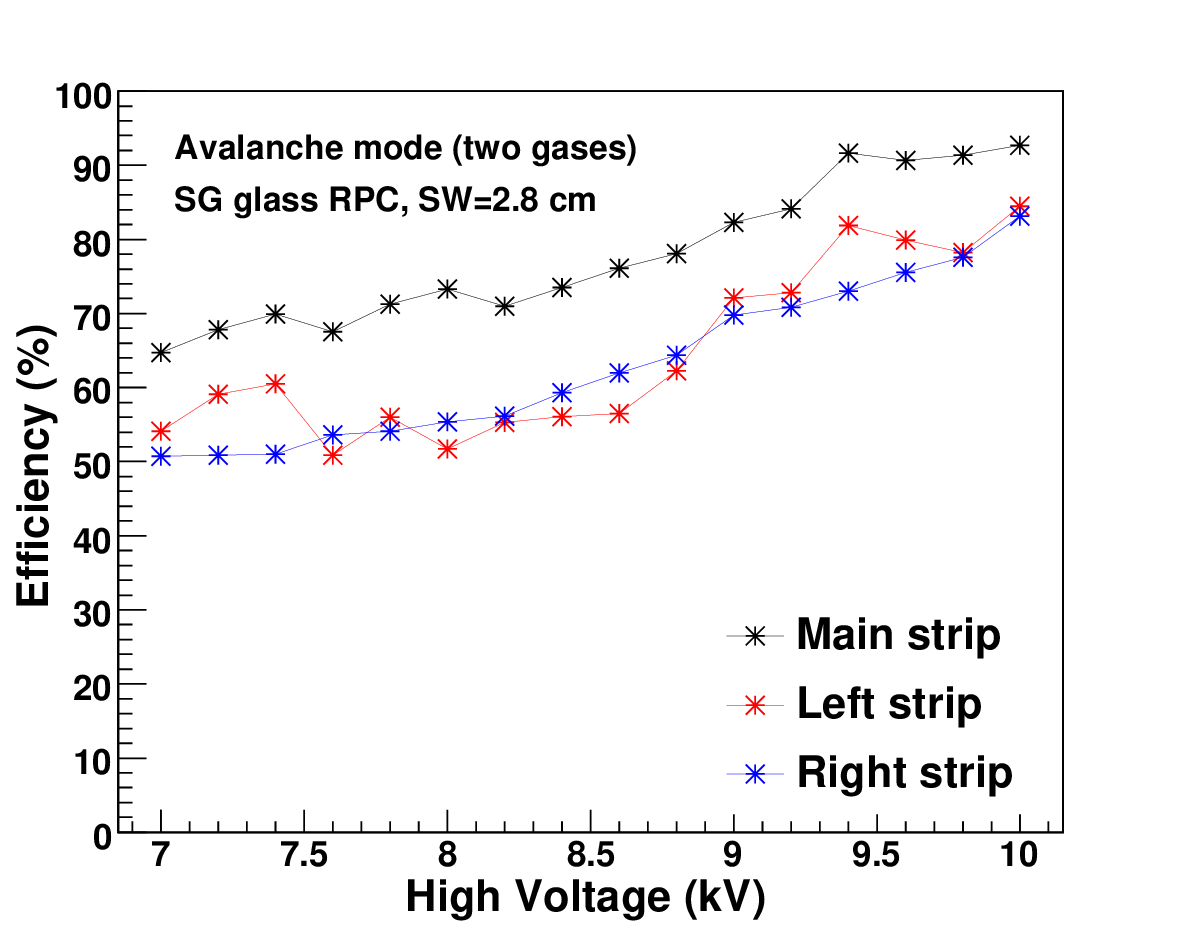}
 \includegraphics[height=.2\textheight]{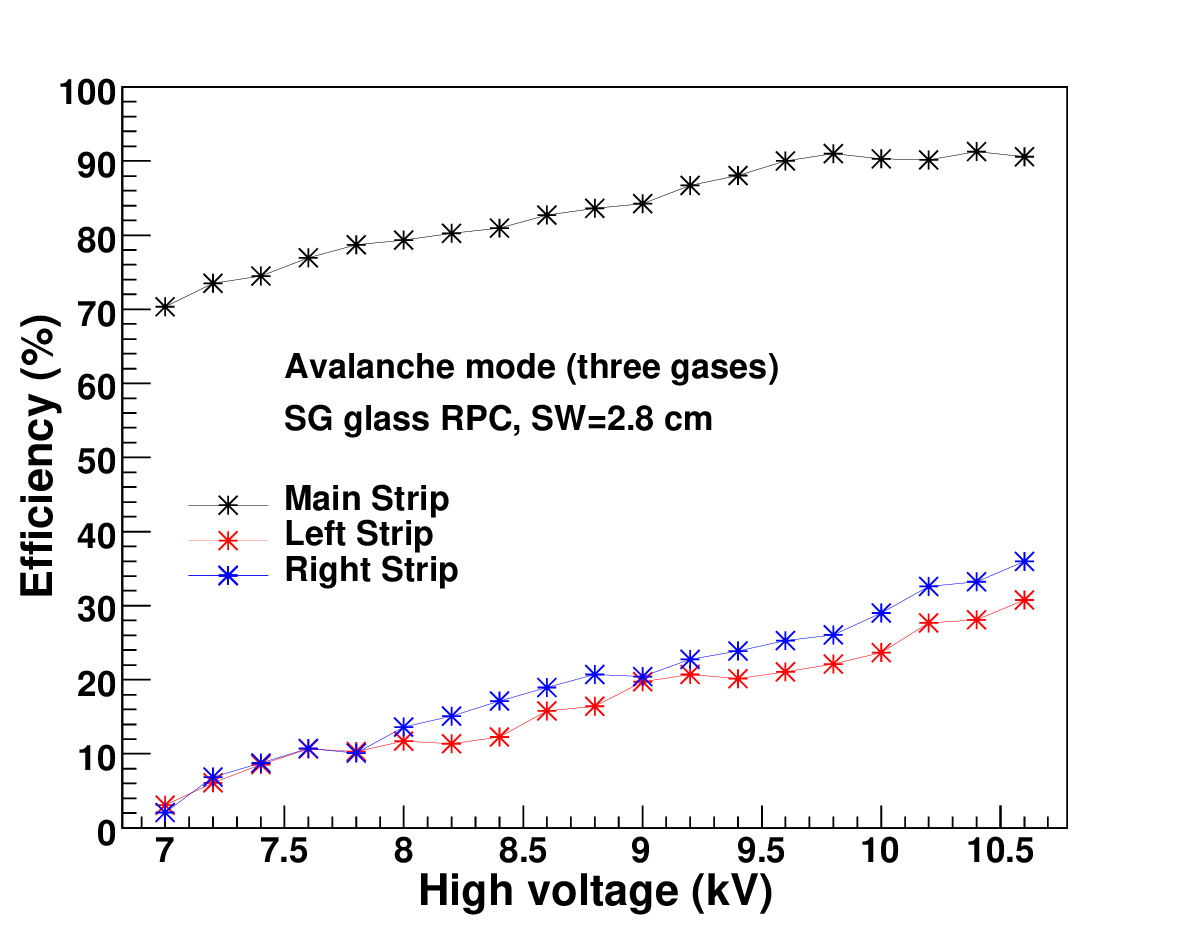}
\includegraphics[height=.2\textheight]{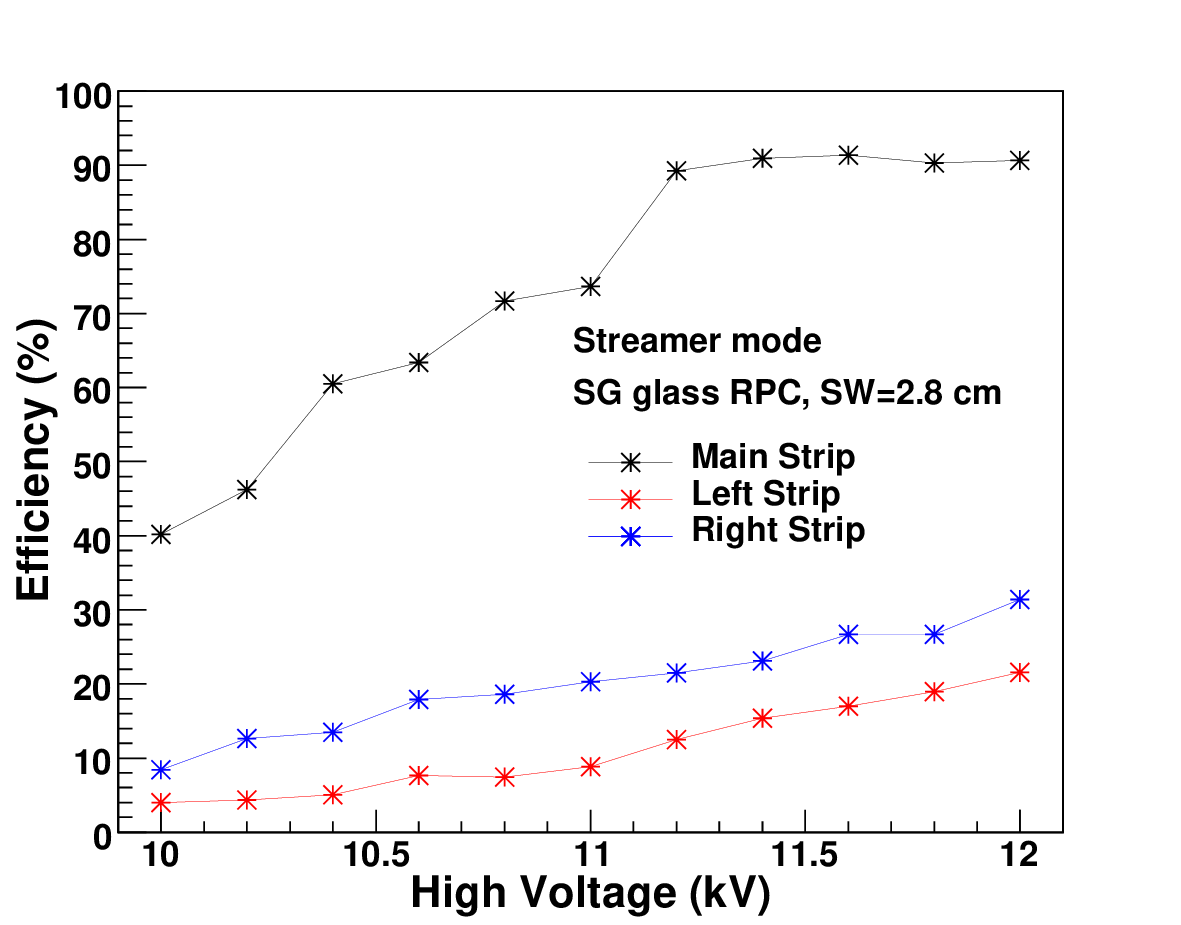}
  \caption{Cross-talk and Efficiency of Saint Gobain glass RPC with the strip width of 2.8 cm in avalanche mode (two gases), avalanche mode (three gases) and streamer mode respectively.}
\label{fig:SG-2.8cm}
\end{figure}

Fig.~\ref{fig:SG-2.8cm} show efficiency and cross-talk of Saint Gobain glass RPC with the strip width of 2.8 cm in avalanche mode (two gases), avalanche mode (three gases) and streamer mode respectively. 

\begin{figure}[htbp]
\renewcommand{\figurename}{Fig.}
\centering
\includegraphics[height=.2\textheight]{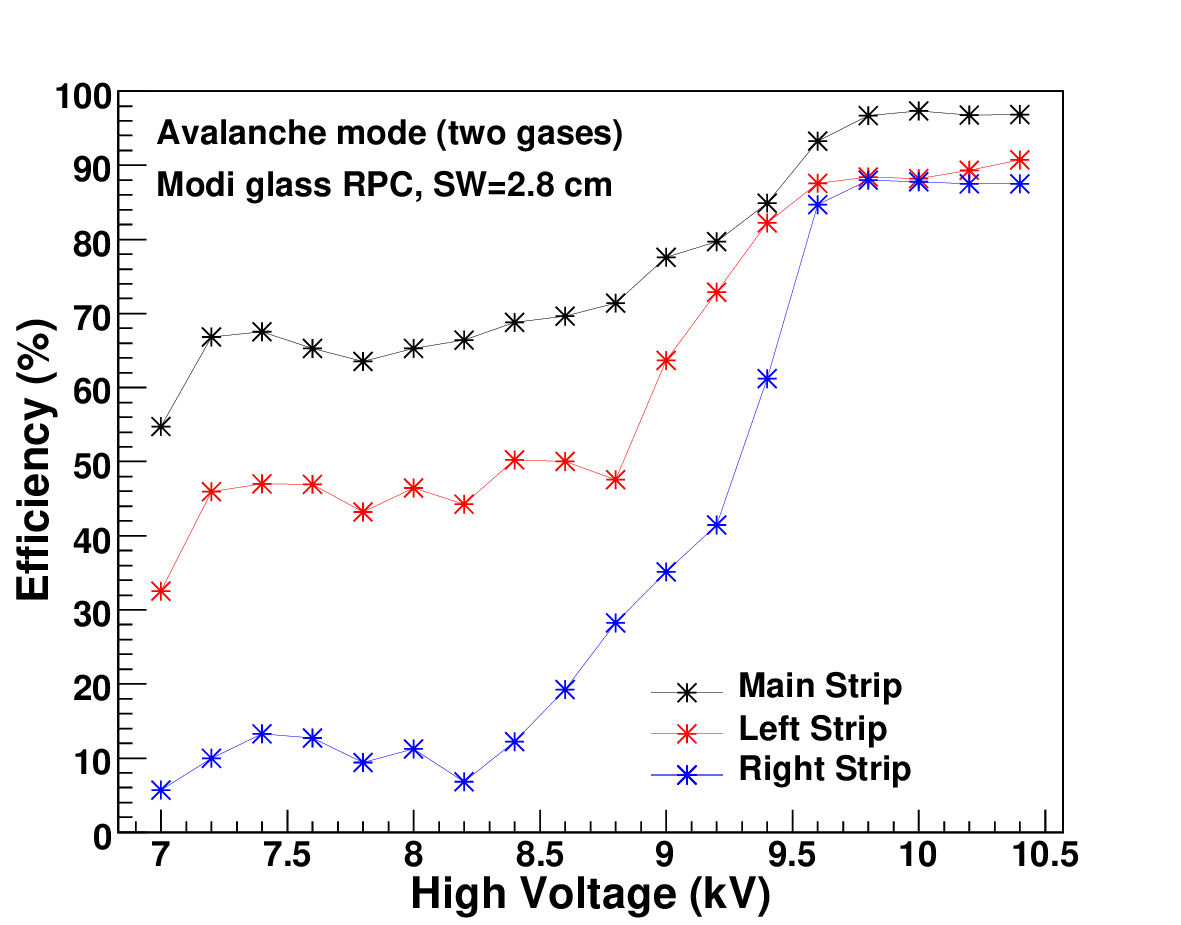}
 \includegraphics[height=.2\textheight]{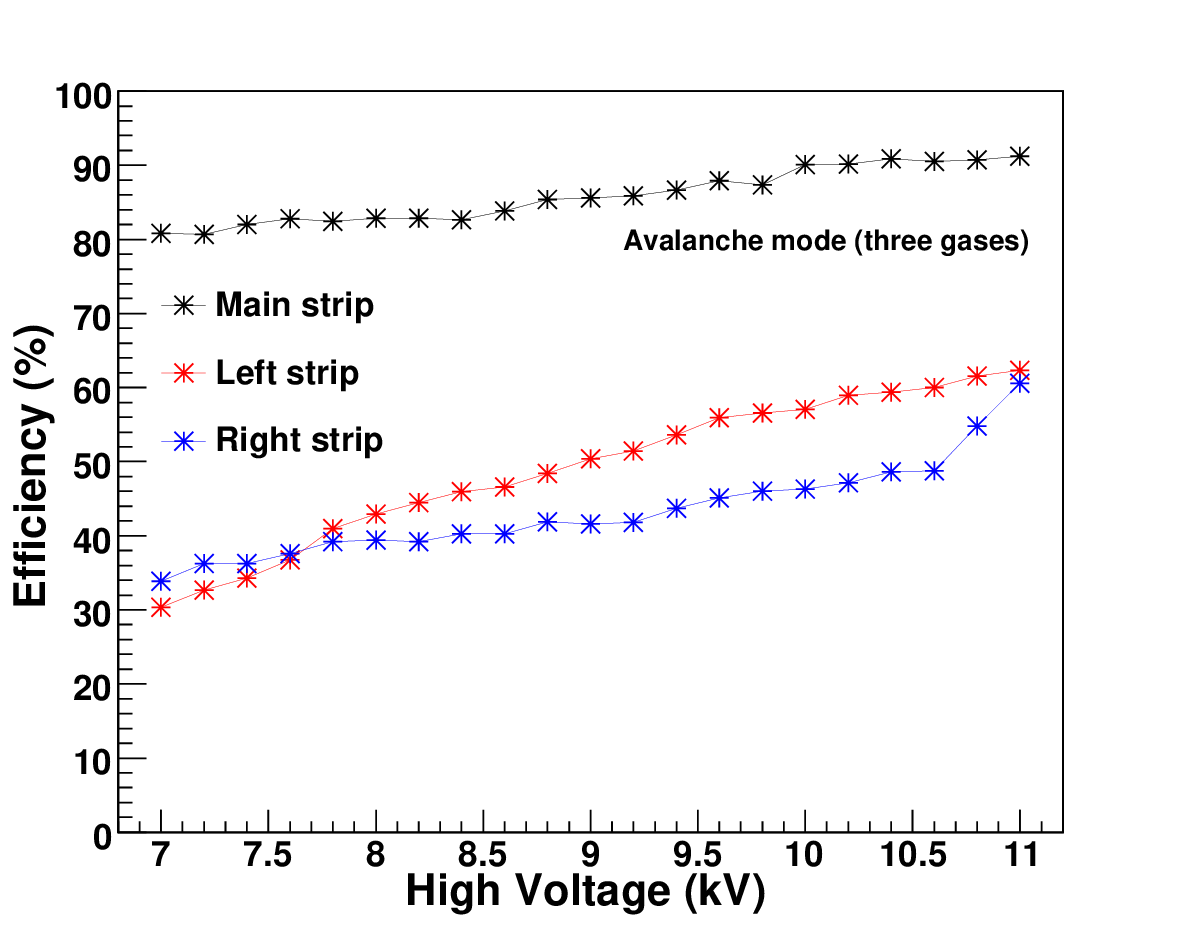}
\includegraphics[height=.2\textheight]{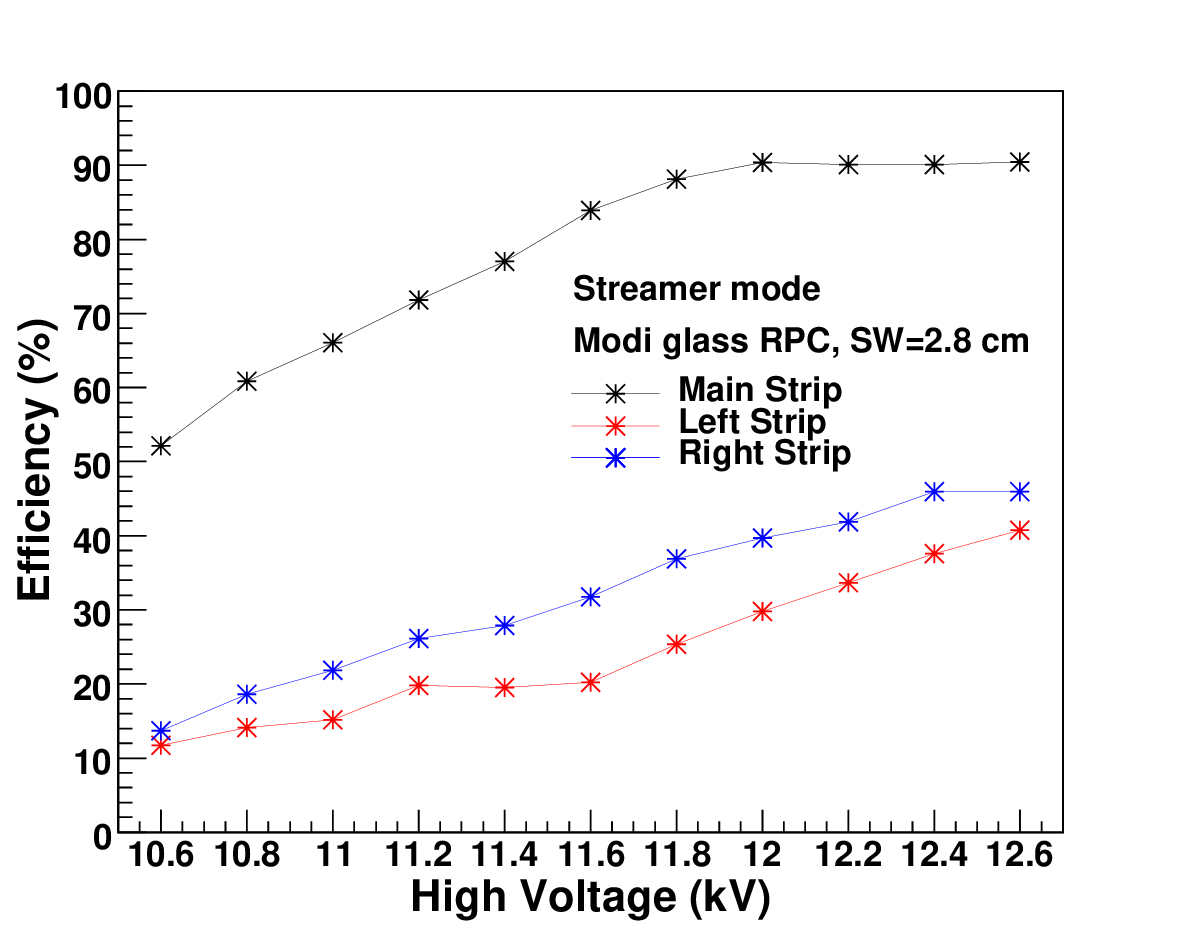}
  \caption{Cross-talk and Efficiency of Modi glass RPC with the strip width of 2.8 cm in avalanche mode (two gases), avalanche mode (three gases) and streamer mode respectively.}
\label{fig:Modi-2.8cm}
\end{figure}

Fig.~\ref{fig:Modi-2.8cm} show efficiency and cross-talk of Modi glass RPC with the strip width of 2.8 cm, in avalanche mode (two gases), avalanche mode (three gases) and streamer mode respectively.

\noindent
Tables~\ref{eff_table} and ~\ref{cross_talk_table} shows efficiency and cross-talk measurements (approximate values) for various glass RPCs operated in different modes at their operating voltages.

\begin{table}[ht]
\small
  \begin{center}
   
    \scalebox{0.9}{
      \begin{tabular}{|c|c|c|c|c|} 
        \hline
        Modes & Strip width (cm) & Asahi (\%) & SG (\%) & Modi (\%) \\ 
        \hline                  
       
        Two gases (avalanche) & & 91 & 92 & 96 \\
        Three gases (avalanche) & 2.8 & 91 & 90 & 91\\
        Streamer & & 90 & 90 & 90 \\
       
        \hline
      \end{tabular}
    }
    \caption{Efficiencies of all the RPCs operated in different modes with different strip widths at their operating voltages.} 
    \label{eff_table} 
  \end{center}
\end{table}

\begin{table}[ht]
\small
  \begin{center}
    
    \scalebox{0.9}{
      \begin{tabular}{|c|c|c|c|c|} 
        \hline 
        Modes & Strip width (cm) & Asahi (\%) & SG (\%) & Modi (\%) \\ 
  
        \hline 
        Two gases (avalanche) & & 84 & 84 & 87 \\
        Three gases (avalanche) & 2.8 & 12 & 30 & 60\\
        Streamer & & 19 & 21 & 43 \\
       
        \hline
      \end{tabular}
    }
    \caption{Cross-talk measurements (approximate values) for all RPCs operated in different modes at their operating voltages.}
    \label{cross_talk_table} 
  \end{center}
\end{table}

\begin{figure}[ht]
\renewcommand{\figurename}{Fig.}
\centering
\includegraphics[height=.2\textheight]{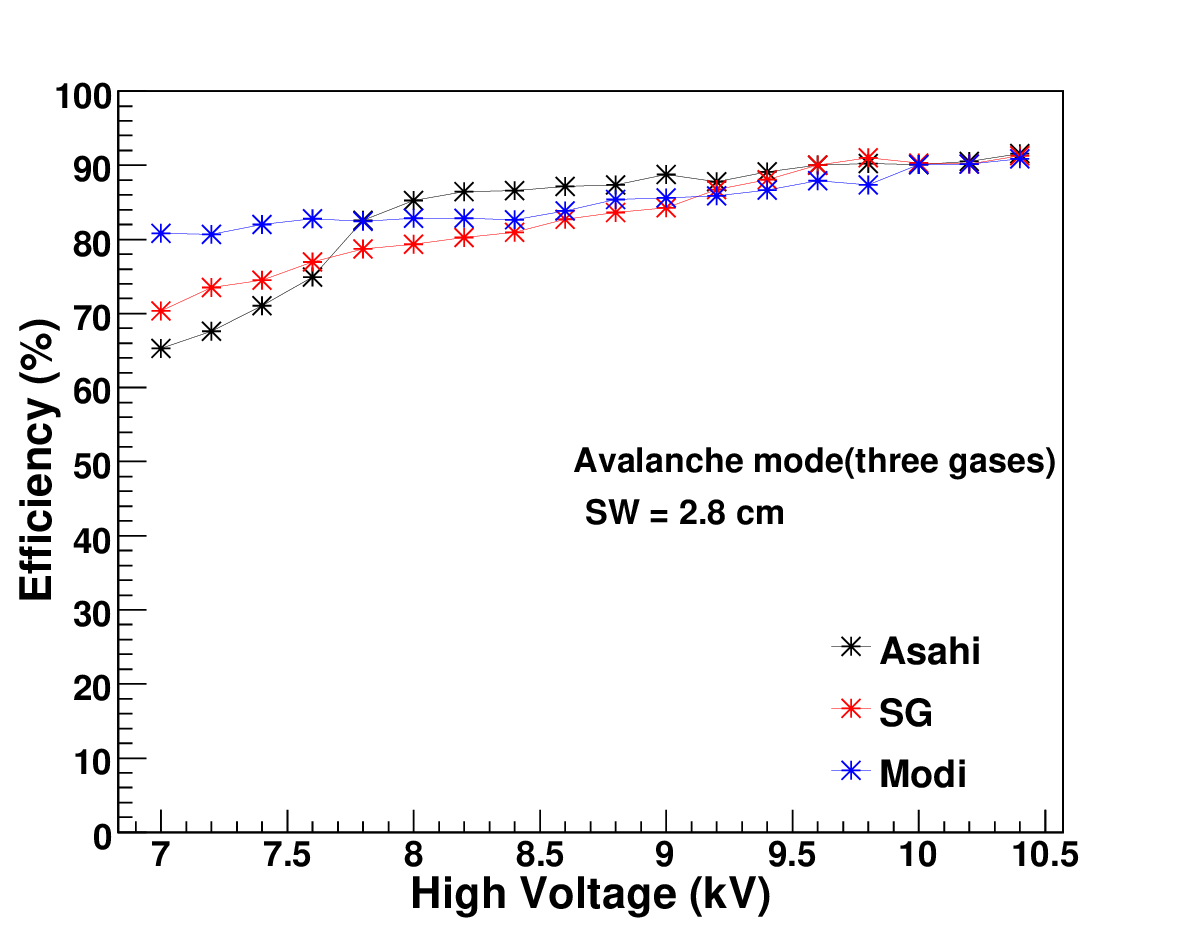}
 \includegraphics[height=.2\textheight]{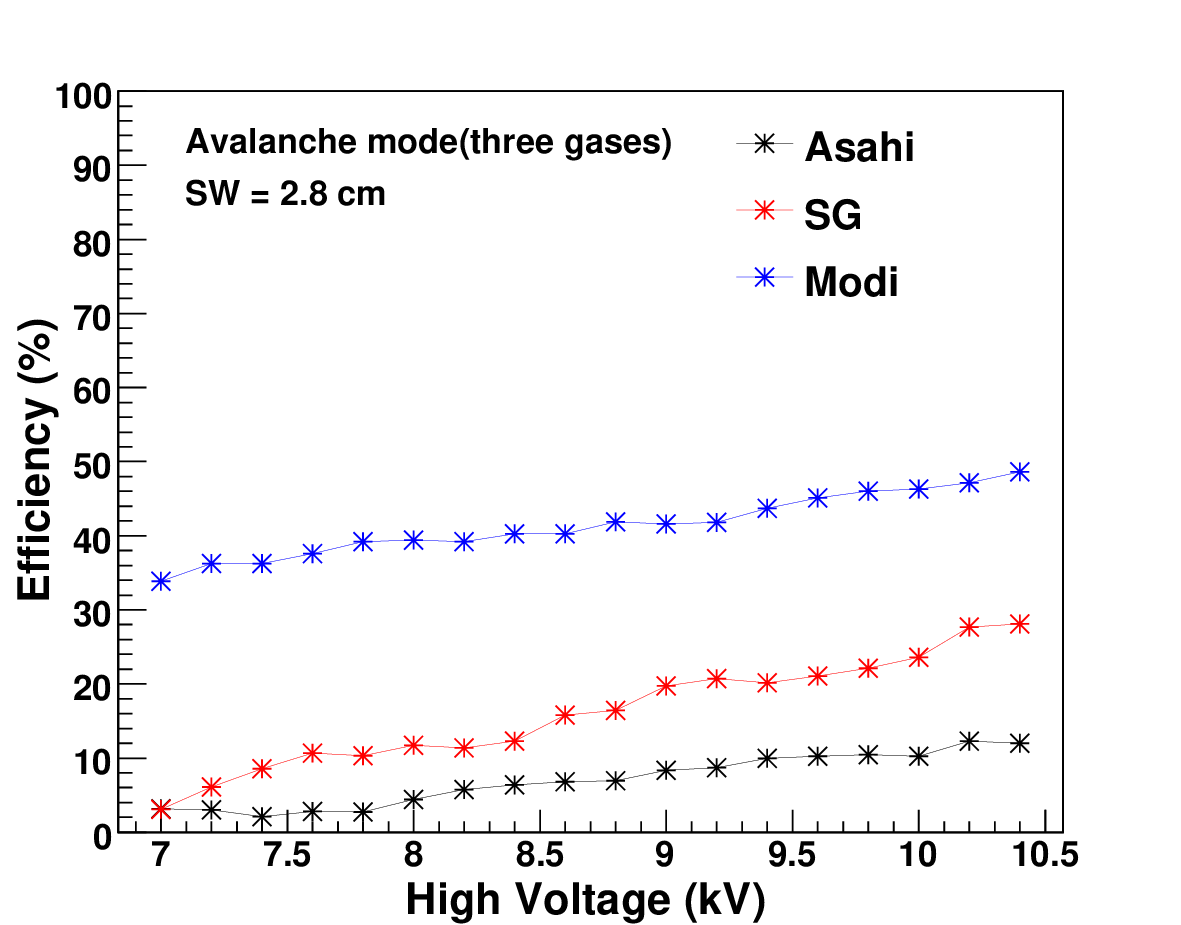}

  \caption{A comparison plot of efficiency (left) and cross-talk (right) of Asahi, Saint Gobain and Modi glass RPC with the strip width of 2.8 cm in avalanche mode (three gases).}
\label{fig:compall-2.8cm}
\end{figure}

It is observed that the higher cross-talk was obtained in case of two gases only (avalanche mode). Also, due to the poor quality of Modi glass RPC a higher cross-talk was observed. It may also be due to the factors like temperature, humidity and noise rate. Cross-talk and Efficiency improves with the addition of $SF_{6}$ and argon gas. A comparison plot of all the RPCs with strip width of 2.8 cm in avalanche mode (three gases) is shown in Fig.~\ref{fig:compall-2.8cm}. 

\section{Discussions and Conclusion}
\label{diss_con}

Resistive plate chambers are the main component of whole ICAL detector and hence a proper R \& D of them is absolutely necessary. Glass is one of the main component of RPC and before making a choice of electrode it needed detailed studies. We procured glass samples of different manufacturers named as Asahi, Saint Gobain and Modi from a local market and compared them on the basis of physical, electrical and optical properties, surface characteristics and elemental composition. We tried to find out a comparative scale, which glass sample is best suited as an electrode in RPCs. On the basis of the properties it was concluded that Asahi and Saint Gobain are better than Modi glass.

We fabricated RPCs of 30 cm $\times$ 30 cm made of the same material which was characterised for the selection of electrode to be used for the RPC. We characterised the three glass RPCs by measuring their cross-talk and efficiency. We conclude that Asahi and Saint Gobain glass RPC gave best results than Modi glass RPC.


\paragraph{Acknowledgements}: We thank CIL department of PU, Cyclotron facility (Nuclear department of PU), NITTTR, Sec 26, Chandigarh for characterisation techniques. We thank engineers and technical staff of PU-EHEP lab. R. Kanishka acknowledges UGC/DAE/DST (Govt. Of India) for funding.


\begin{thebibliography}{99}

\bibitem{kajita} Takaaki Kajita, Arthur B. McDonald, {\it Metamorphosis in the particle world}, The Nobel Prize in Physics 2015, \href{http://www.nobelprize.org/nobel\_prizes/physics/laureates/2015/press.pdf}{http://www.nobelprize.org/nobel\_prizes/physics/laureates/2015/press.pdf}, 6 October 2015.

\bibitem{ino:2006} M. S. Athar et al., {\it India-based Neutrino Observatory: Project Report Volume I}, \href{http://www.ino.tifr.res.in/ino/OpenReports/INOReport.pdf}{http://www.ino.tifr.res.in/ino/OpenReports/INOReport.pdf} (2006).

\bibitem{white} R. N. Mohapatra et al., {\it Theory of neutrinos: A White Paper}, \href{http://arxiv.org/abs/hep-ph/0510213v2}{arXiv:hep-ph/0510213v2}, 2 Dec 2005.

\bibitem{indu:2006} D. Indumathi et al., {\it Neutrino oscillation probabilities: Sensitivity to parameters}, \href{http://arxiv.org/abs/hep-ph/0603264v2}{arXiv:hep-ph/0603264v2}, 13 April 2006.

\bibitem{physics} T.~Thakore, A.~Ghosh, S.~Choubey and A.~Dighe, {\it The Reach of INO for Atmospheric Neutrino Oscillation Parameters}, JHEP {\bf 1305}, 058, \href{http://arxiv.org/abs/1303.2534}{arXiv:hep-ph/1303.2534} (2013).

\bibitem{physics1} A.~Ghosh, T.~Thakore and S.~Choubey, {\it Determining the Neutrino Mass Hierarchy with INO, T2K, NOvA and Reactor Experiments}, JHEP {\bf 1304}, 009, \href{http://arxiv.org/abs/1212.1305}{arXiv:hep-ph/1212.1305} (2013).

\bibitem{satya} B. Satyanarayana, {\it Design and Characterisation Studies of Resistive Plate Chambers}, Ph.D thesis, Department of Physics, IIT Bombay, PHY-PHD-10-701, (2009).

\bibitem{central} A. Chatterjee et al., {\it A Simulations Study of the Muon Response of the Iron Calorimeter detector at the India-based Neutrino Observatory}, JINST {\bf 9} P07001, \href{http://arxiv.org/abs/1405.7243}{[arXiv:1405.7243]} (2014).

\bibitem{hadrons} M. M. Devi et al., {\it Hadron energy response of the Iron Calorimeter detector at the India-based Neutrino Observatory}, JINST {\bf 8} P11003, \href{http://arxiv.org/abs/1304.5115}{[arXiv:1304.5115]} (2013).

\bibitem{kolahal} Kolahal Bhattacharya, et al., (INO Collaboration),
{\it Error propagation of the track model and track fitting strategy for the iron calorimeter detector in India-based neutrino observatory}, Computer Physics Communications (Elsevier) {\bf 185} (12) 3259--3268, (2014).

\bibitem{Santonico} R. Santonico, R. Cardarelli, Nucl. Instrum. Meth. A 187, 377 (1981).

\bibitem{Santonico1} R. Santonico, R. Cardarelli, Nucl. Instrum. Meth. A 263, 20 (1988).

\bibitem{belle} Abe T. et al., (Belle-II Collaboration), Belle II Technical Design Report, \href{http://arxiv.org/abs/1011.0352} {arXiv:physics.ins-det/1011.0352}, KEK-REPORT-2010-1.

\bibitem{cms} CMS Collaboration, Technical Proposal, CERN-LHCC-94-38, 1994; CMS Collaboration, JINST 3, S08004 (2008).

\bibitem{kalmani:2009} S. D. Kalmani et al., {\it On-line gas mixing and multi-channel distribution system}, Nucl. Instrum. Meth. A (602), 2009.

\bibitem{carda1} R. Cardarelli et al., Nucl. Instrum. Meth. A 382, 470 (1996).

\bibitem{carda} R. Cardarelli et al., Nucl. Instrum. Meth. A 333, 399 (1993).

\bibitem{fonte} P. Fonte, Nucl. Instrum. Meth. A 456, 6 (2000).

\bibitem{fonte1} P. Fonte, IEEE Trans. on Nucl. Sci. 49, 881 (2002).

\bibitem{aging} B. Satyanarayana et al., Nucl. Instrum. Meth. B 158, 195--198 (2006). 

\bibitem{Czyrkowski} H. Czyrkowski, et al., Nucl. Instrum. Meth. A 419, 490 (1998).

\bibitem{Meghna} K. K. Meghna, et al., Journal of Instrumentation 7, P10003 (2012).

\bibitem{Bhide} Sarika Bhide, V. M. Datar, Satyajit Jena, S. D. Kalmani, N. K. Mondal, G. K. Padmashree, B. Satyanarayan, R. R. Shinde, P. Verma, Pramana - Journal of Physics 69 (6), 1015 (2007).

\bibitem{Datar} V. M. Datar, Satyajit Jena, S. D. Kalmani, N. K. Mondal, P. Nagaraj, L. V. Reddy, M. Saraf, B. Satyanarayan, R. R. Shinde, P. Verma, Nucl. Instrum. Meth. A 602, 744 (2009).

\bibitem{strip_kan} R. Kanishka et al., {\it Strip Width Optimisation of RPC Detectors for India-based Neutrino Observatory}, manuscript in preparation.

\bibitem{Mengucci} A. Mengucci, A. Paoloni, M. Spinetti, L. Votano, Nucl. Instrum. Meth. A 583, 264 (2007).

\bibitem{Cwoik} M. Cwoik, W. Dominik, M. Gorski, J. Krolikowski, Nucl. Instrum. Meth. A 508, 38 (2003).

\bibitem{Lie-Hua} M. A. Lie-Hua, et al., Chinese Physics C 34 (8), 1116 (2010).

\bibitem{Salim} M. Salim, R. Hasan, N. Majumdar, S. Mukhopadhayay, B. Satyanarayana, JINST 7, P11019 (2012).

\bibitem{Abe} K. Abe, et al., Nucl. Instrum. Meth. A 455, 397 (2000).

\bibitem{bueno} C. C. Bueno et al., Proc. Intl. Nucl. Atlantic Conf., Brazil, (2007).

\bibitem{abe} K. Abe et al., Nucl. Instrum. Meth. A 455, 397 (2000).




\end{thebibliography}
\end{document}